\documentclass[useAMS,usenatbib]{mn2e}
\usepackage{rotating}
\usepackage{amssymb} 
\usepackage{amsmath} 
\usepackage{graphicx}

\author[Noutsos et al.]
{A. Noutsos$^1$, S. Johnston$^2$, M. Kramer$^1$ and A. Karastergiou$^3$\\
$^1$ University of Manchester, Jodrell Bank Observatory, Macclesfield, 
Cheshire, SK11 9DL.\\
$^2$ Australia Telescope National Facility, CSIRO, P.O. Box 76, 
Epping, NSW 1710, Australia.\\
$^3$ Astrophysics, University of Oxford, Denys Wilkinson Building, Keble Road,
Oxford OX1 3RH, UK\\
} 
\date{\today} 
\title[New pulsar rotation measures]
{New pulsar rotation measures and the Galactic magnetic field}

\def\jnl@style{\rm}
\def\aaref@jnl#1{{\jnl@style#1}}

\def\aaref@jnl#1{{\jnl@style#1\thinspace}}

\def\aj{\aaref@jnl{{\em AJ}}}                   
\def\araa{\aaref@jnl{{\em ARA\&A}}}             
\def\apj{\aaref@jnl{{\em ApJ}}}                 
\def\apjl{\aaref@jnl{{\em ApJ}}}                
\def\apjs{\aaref@jnl{{\em ApJS}}}               
\def\ao{\aaref@jnl{{\em Appl.~Opt.}}}           
\def\apss{\aaref@jnl{{\em Ap\&SS}}}             
\def\aap{\aaref@jnl{{\em A\&A}}}                
\def\aapr{\aaref@jnl{{\em A\&A~Rev.}}}          
\def\aaps{\aaref@jnl{{\em A\&AS}}}              
\def\azh{\aaref@jnl{{\em AZh}}}                 
\def\baas{\aaref@jnl{{\em BAAS}}}               
\def\jrasc{\aaref@jnl{{\em JRASC}}}             
\def\memras{\aaref@jnl{{\em MmRAS}}}            
\def\mnras{\aaref@jnl{{\em MNRAS}}}             
\def\pra{\aaref@jnl{{\em Phys.~Rev.~A}}}        
\def\prb{\aaref@jnl{{\em Phys.~Rev.~B}}}        
\def\prc{\aaref@jnl{{\em Phys.~Rev.~C}}}        
\def\prd{\aaref@jnl{{\em Phys.~Rev.~D}}}        
\def\pre{\aaref@jnl{{\em Phys.~Rev.~E}}}        
\def\prl{\aaref@jnl{{\em Phys.~Rev.~Lett.}}}    
\def\pasp{\aaref@jnl{{\em PASP}}}               
\def\pasj{\aaref@jnl{{\em PASJ}}}               
\def\qjras{\aaref@jnl{{\em QJRAS}}}             
\def\skytel{\aaref@jnl{{\em S\&T}}}             
\def\solphys{\aaref@jnl{{\em Sol.~Phys.}}}      
\def\sovast{\aaref@jnl{{\em Soviet~Ast.}}}      
\def\ssr{\aaref@jnl{{\em Space~Sci.~Rev.}}}     
\def\zap{\aaref@jnl{{\em ZAp}}}                 
\def\nat{\aaref@jnl{{\em Nature}}}              
\def\iaucirc{\aaref@jnl{{\em IAU~Circ.}}}       
\def\aplett{\aaref@jnl{{\em Astrophys.~Lett.}}} 
\def\apspr{\aaref@jnl{{\em Astrophys.~Space~Phys.~Res.}}}
\def\bain{\aaref@jnl{{\em Bull.~Astron.~Inst.~Netherlands}}} 
\def\fcp{\aaref@jnl{{\em Fund.~Cosmic~Phys.}}}  
\def\gca{\aaref@jnl{{\em Geochim.~Cosmochim.~Acta}}}   
\def\grl{\aaref@jnl{{\em Geophys.~Res.~Lett.}}} 
\def\jcp{\aaref@jnl{{\em J.~Chem.~Phys.}}}      
\def\jgr{\aaref@jnl{{\em J.~Geophys.~Res.}}}    
\def\jqsrt{\aaref@jnl{{\em J.~Quant.~Spec.~Radiat.~Transf.}}}
\def\memsai{\aaref@jnl{{\em Mem.~Soc.~Astron.~Italiana}}}
\def\nphysa{\aaref@jnl{{\em Nucl.~Phys.~A}}}   
\def\physrep{\aaref@jnl{{\em Phys.~Rep.}}}   
\def\physscr{\aaref@jnl{{\em Phys.~Scr}}}   
\def\planss{\aaref@jnl{{\em Planet.~Space~Sci.}}}   
\def\procspie{\aaref@jnl{{\em Proc.~SPIE}}}   

\makeatletter
\DeclareRobustCommand{\Cpp}
{\valign{\vfil\hbox{##}\vfil\cr
   \textsf{C\kern-.1em}\cr
   $\hbox{\fontsize{\sf@size}{0}\textbf{+\kern-0.05em+}}$\cr}%
}
\makeatother

\begin{document}

\bibliographystyle{mn2e}

\maketitle

\begin{abstract} We measured a sample of 150 pulsar Rotation Measures (RMs) 
  using the 20-cm receiver of the Parkes 64-m radio telescope. 46 of
  the pulsars in our sample have not had their RM values previously published,
  whereas 104 pulsar RMs have been revised. We used a novel quadratic fitting
  algorithm to obtain an accurate RM from the calibrated polarisation
  profiles recorded across 256 MHz of receiver bandwidth. The new data are
  used in conjunction with previously known Dispersion Measures (DMs) and the
  NE2001 electron-density model to study models of the direction and magnitude
  of the Galactic magnetic field.
\end{abstract}

\section{Introduction}
Magnetic fields are an integral part of nature. No physical
description can be complete without considering them, whether one
deals with the realm of the very small (e.g.~quantised Hall effect,
\nocite{kli86}von Klitzing 1986) or that of the very large (e.g.~cosmological magnetic
fields, \nocite{zel65}Zel'Dovich 1965). Hence, the study of magnetic fields is
essential in understanding the universe. The birth of radio-astronomy
in the 1930s allowed, amongst others, the measurement of galactic
magnetic fields. The closest and most relevant to us galactic magnetic
field is, of course, that of the Milky Way.

The large-scale component of the Galactic magnetic field (hereafter
GMF) is intertwined with a number of interesting astrophysical
processes, such as the deflection of ultra-high-energy cosmic rays
(UHECRs) by the GMF (e.g.~\nocite{mdh98}Medina Tanco et al. 1998), or the existence of
star-forming regions (e.g.~\nocite{nkl+06}Novak et al. 1998). A number of observables
have been employed towards mapping the GMF: e.g.~Zeeman splitting of
spectral lines, optical starlight polarisation and Faraday rotation of
polarised radio sources. However, the main challenge that most of
these methods face arises from the observer's location inside
the field's physical boundaries. Faraday rotation of polarised sources
exploits this fact by providing a measure of the strength and
direction of the line-of-sight ({\em los}) component of the GMF. As was first
pointed out in \citet{ls68}, Faraday rotation of radio pulsars, in
particular, is a powerful method for mapping the GMF. One can summarise
the value of pulsars towards this goal as follows: (a) pulsars are
distributed throughout the entire volume of the Galactic disc; (b) they are amongst the most
polarised radio sources known; and, finally, (c) the dispersion of
their pulses, given by their dispersion measure (DM), not only allows
one to estimate their distance to a high accuracy, but also, when
combined with their rotation measure (RM), provides a direct way of
estimating the average value of the interstellar magnetic field.

The potential of Faraday rotation as a probe to the GMF has motivated
campaigns to try and measure pulsar RMs. The first such efforts, by
\citet{man72} and \citet{man74}, brought the number of determined
RMs to 38. An important contribution 
was made by \citet{hl87}, who measured 163 values, increasing the
total number of available RMs to 185. Subsequent observations by
\citet{cmh91}, \citet{rl94}, \citet{qmlg95}, \citet{vdhm97} and
\citet{hmq99} brought the total to 320. To date, the pulsar catalogue
\nocite{mhth05} contains 554 RMs: i.e.~nearly a third of the pulsars
known have an RM value.

The first use of available pulsar polarisation data to map the GMF led
to unequivocal evidence for a clockwise (CW) directed local field, as viewed from the Galactic north, of uniform
strength $\sim 2$ $\mu$G (\nocite{man72}Manchester 1972; \nocite{man74}Manchester 1974). Re-analysis
of 48 pulsar RMs from the \citet{mt77} catalogue provided the first
indication for a magnetic-field reversal in the Carina--Sagittarius
arm \citep{tn80}. These finds were confirmed by the work of
\citet{ls89}, who also suggested an additional reversal outside the
Perseus arm. As more data were being collected, the field's structure
to larger distances could be explored: e.g.~\citet{rl94} and
\citet{hmlq02} reported magnetic-field reversals in the Crux-Scutum and
Norma regions.

Despite the wealth of information that observational data can provide,
the sole use of pulsar RMs and DMs for the estimation of the
interstelar field cannot provide sufficient information about the GMF
structure: this is because the quantity RM/DM can
only provide a weighted average of the parallel component of the
field, with the averaging performed across the distance between
pulsars. But since both the electron density and the magnetic field
are far from simple functions of azimuth and distance, inverting
RM/DM to obtain $\boldsymbol{B}$ is not feasible.

A possible solution is to try and fit magnetic-field models to the
data and find which model is more representative of the observed
RMs (see section 6). A number of authors have tried various field configurations to
match the available data: e.g. \citet{sf83} were amongst the first to
favour a 2-arm bisymmetric-spiral model, whereas \citet{rk89},
\citet{rl94} and \citet{val05} fitted a concentric ring model: the
latter type of model showed consistency with earlier data, but more
recent observations by \citet{hml+06} contradict the model
predictions, especially towards the inner Galaxy.

Despite those efforts, however, the shape of the GMF remains
unclear. The main difficulties in modelling arise from (a) a
short-scale, seemingly random magnetic-field component --- of the same
order of magnitude as the large-scale one --- that causes large RM
fluctuations across the entire Galactic plane (hereafter GP); (b) regional anomalies
of the ISM (e.g.~the Gum Nebula, the North Polar Spur), which corrupt
a potentially smooth RM variation; and, in addition, (c) the pulsar
distances --- upon which a large portion of modelling is relied ---
are not certain enough, which causes discontinuities in the field
direction over large longitude ranges. 

In this and forthcoming publications we will study the difficulties of
modelling the GMF in some detail. In this paper we will concentrate first on
the measurement of new, accurate RMs in order to supplement previously
sparsely sampled regions of the Galaxy. We will then examine as to whether
commonly studied models of GMF are consistent with the new constraints
provided by our often more precise measurements, and how they can be improved
to provide a better fit to the data. A full modelling of the GMF with the full
sample of available data will be undertaken in a later publication.

\section{Observations} 
We observed a sample of 239 pulsars with the 20-cm H--OH
receiver of the Parkes 64-m radio telescope in order to obtain
a polarization database at high time resolution for all these objects.
The observing session was carried out from 2006 August 24 to 27. The
H--OH receiver at Parkes is equipped with two, orthogonal linear feeds
that are sensitive to polarised emission with an equivalent system
flux density of 43 Jy (\nocite{joh02}Johnston 2002). The full Stokes profiles of each
pulsar were recorded using digital correlator back end, across
256 MHz of bandwidth split in 1024 frequency channels. The wide-band
correlator is capable of producing high-resolution profiles with 1024
phase bins, and more than 60\% of the profiles were recorded in this
mode, while the rest had 512-bin resolution. We observed each of the 239 pulsars
for $\sim$15 min, producing $\sim$90 10-s
subintegrations. In order to save disk space, we further time-averaged 
the data to reduce them to 15 1-min subintegrations, which we later calibrated and analysed.

\section{Data Analysis} 
We processed all the data using the PSRCHIVE software
package \citep{hvm04}, within which we developed new software to perform
RM fitting. The raw data from the wide-band correlator
first underwent a process of RFI excision and polarisation
calibration. Following this, each data set was summed in time to obtain 
four integrated Stokes profiles for each frequency channel. 
One average polarisation position angle
(PA) was computed per frequency channel from the Stokes
parameters $Q$ and $U$, summed across $n$ phase bins $i$ corresponding
to the pulse, as:
\begin{equation} 
\label{eq:ppaeq} 
{\rm PA} = \frac{1}{2}\arctan{\left(\frac{\sum_{i=0}^{n}U_i}{\sum_{i=0}^{n}Q_i}\right)}
\end{equation} 

We computed the error of this average PA using one of the two
following methods, depending on the signal-to-noise ratio of the total
linear polarization $L=({\sum_{i=0}^nQ^2+\sum_{i=0}^nU^2})^{1/2}$, given
by ${\rm s/n}=(1/\sqrt{n})L/\sigma_I$, where $\sigma_I$ is the RMS of the
total intensity. For low s/n ratios ($<10$), the 1$\sigma$ error on
the PAs, $\sigma_{\rm PA}$, was computed numerically using the method
described in \citet{nc93}.  For high s/n PAs, i.e. $\geq 10$, the
PA error was computed using the formula from \citet{ew01}:
\begin{equation} 
\label{eq:ppaerreq3} 
\sigma_{\rm PA}=\frac{\sqrt{n}}{2}\frac{\sigma_I}{L}
\end{equation}

In obtaining the average PA across the pulse per frequency channel as
described above, there is a possibility that the integration of $Q$
and $U$ may in fact lead to a low average linear polarization and,
therefore, a large error on the PA. This depends on the shape of the
PA profile; integration of consecutive bins with orthogonal PAs
significantly reduces the total linear polarization. While this
possibility exists, we found that in our data the error on the average
PA was significantly improved with respect to the errors on the PA in
individual bins. These reduced errors were instrumental to good RM
fits, therefore we deemed the process of averaging the PA appropriate.

\subsection{The Rotation Measure from a Quadratic Fit with Wraps} 
The degree of Faraday rotation, $\Delta{\rm PA}$, that electromagnetic waves of
wavelength $\lambda$ undergo as they propagate through the interstellar medium
from the pulsar to the Earth is expressed by the Rotation Measure (RM). This
rotation is caused by the interaction of the radio waves with the magnetised
plasma of the interstellar medium and, more specifically, the plasma electrons
along the line-of-sight ({\em los}) to the pulsar. If $n_e$ is the number
density of the plasma electrons along a unit column of length equal to the
distance to the pulsar, $d$, and $\boldsymbol{B}$, the magnetic field of that
plasma, then RM is given as the path integral of $n_e\boldsymbol{B}$ along the
line of sight:
\begin{equation}
\label{eq:rmeq1} 
{\rm RM} = \frac{e^3}{2\pi m_e^2 c^4}\int_{0}^{d}n_e(s)\boldsymbol{B}(s)\cdot d\boldsymbol{s} 
\end{equation} 
where $d\boldsymbol{s}$ is the path vector element in the direction of wave propagation,
$e$ is the electron charge, $m_e$, the electron mass and $c$, the speed of light.
Using cgs units for the constants and expressing the distance in pc and $B$ in $\mu$G, we get

\begin{equation}
\label{eq:rmeq11} 
{\rm RM} = 0.812 \  \int_{0}^{d}\left[\frac{n_e(s)}{{\rm cm}^{-3}}\right]\left[\frac{\boldsymbol{B}(s)}{\mu {\rm G}}\right]\cdot \left(\frac{d\boldsymbol{s}}{{\rm pc}}\right) 
\end{equation} 
so that RM is expressed in the usual units of rad m$^{-2}$.

An observable property of the above magneto-optical interaction is
that, for a particular pulsar, the longer the wave, the greater the
effect of Faraday rotation, with infinitely short waves being
unaffected. Mathematically, this realtionship is expressed as
\begin{equation} 
\label{eq:rmeq2} 
\Delta{\rm PA}={\rm RM} \lambda^2=c^2{\rm RM}\frac{1}{f^2} 
\end{equation} 
where $f$ is the frequency of the waves.

Eq.~\ref{eq:rmeq2} implies that the RM can be measured across the
observation band by fitting a quadratic function to the PAs of every
frequency channel. Using a Bayesian approach, we developed a fitting algorithm
that was applied to the available data sets. One of the main
strengths of the quadratic fitting algorithm is that it accounts for
the $180^\circ$ ambiguity that the PAs are subject to according to
Eq.~\ref{eq:ppaeq}. The algorithm finds the best fit to the data by
means of a Bayesian likelihood test. The fitting function was of the
form, ${\rm PA}={\rm PA}_0 + c^2{\rm RM}(1/f_j^2-1/f_0^2)$, where ${\rm PA}_0$ is the PA
at frequency $f_0$, and $f_j$ is the frequency of channel $j$. The
free parameters ${\rm PA}_0$ and RM were stepped through the ranges
$[0,\pi]$ rad and $[-1000,1000]$ rad m$^{-2}$, in $1^\circ$ and 1--rad
m$^{-2}$ steps, respectively.

\begin{figure}
\vspace*{10pt} 
\includegraphics[width=0.47\textwidth]{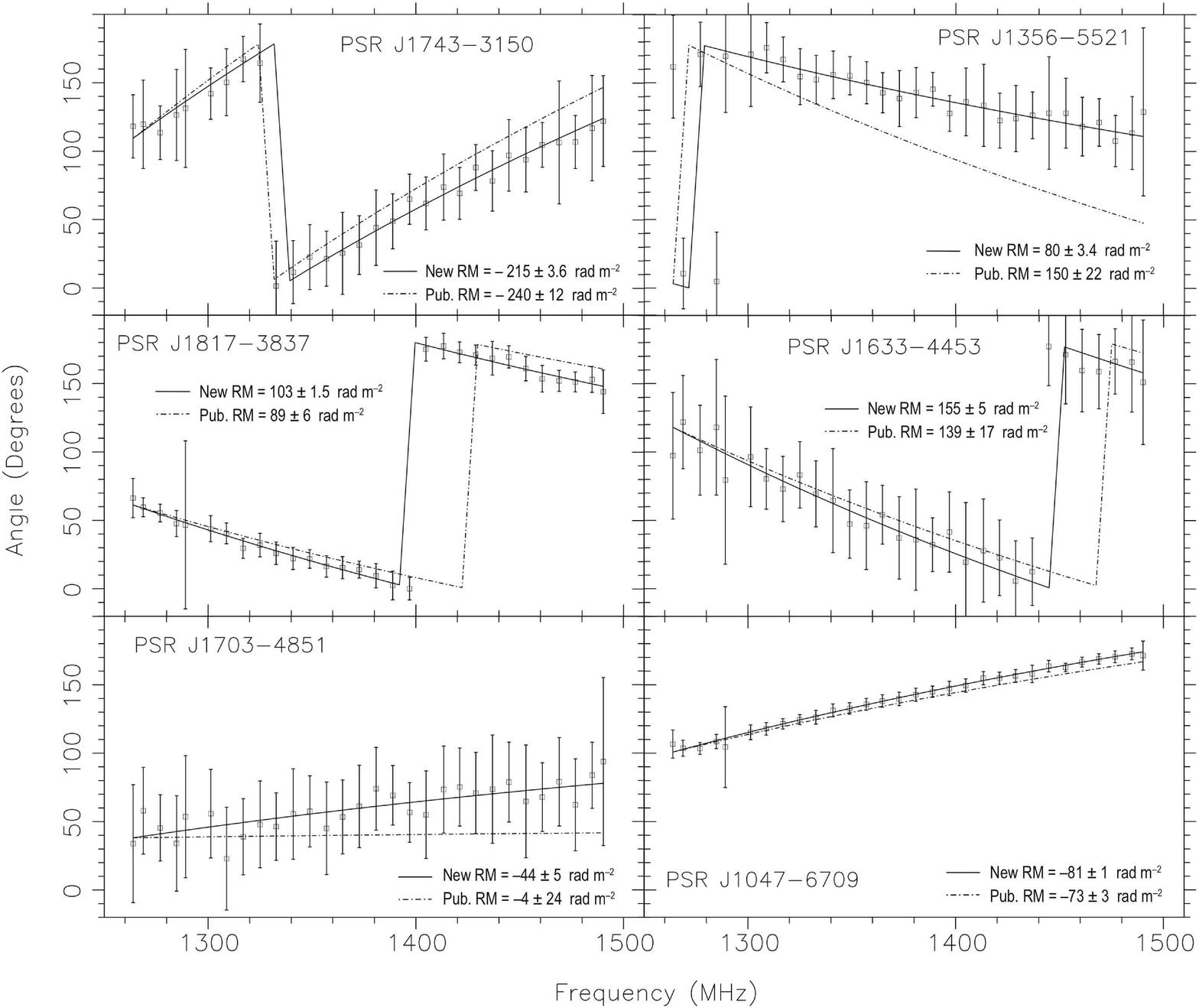} 
\caption{\label{fig:quadFitExamples} Examples of RM results obtained
using a quadratic fit with wraps. The position angle, PA, is plotted
against frequency.  The solid lines represent the best fit from the
method described here, whereas the dashed--dotted lines correspond to
the fits based on previously published RMs by \citet{hml+06}.}
\end{figure}

The accuracy of the method is clearly displayed in the examples of
Fig.~\ref{fig:quadFitExamples}. Our confidence in the calculated RMs is drawn both from the
well calibrated data and the robust fitting approach, but also from the meticulous determination of the RM errors.
The latter, we believe, has produced more firm results compared to previous efforts. 
The following two sections describe the procedure for the calculation of these errors.  

\subsection{Statistical Error on the RM from Monte Carlo Simulations} 
In order to estimate the statistical error on the best-fit RMs, $\sigma_{\rm stat}$, we 
generated a look-up table of RM errors corresponding to the typical (average) PA error in our data, $\sigma_{\rm PA}^{\rm data}$,
and the actual RM value that came out of the fit to the data, ${\rm RM}^{\rm data}$. 

In order to create the look-up table, we generated Monte Carlo (MC) data sets that contained fake PA values
which were calculated from a grid of chosen RM values, $\langle{\rm RM}^{\rm mc}\rangle$, using Eq.~\ref{eq:rmeq2}.
The observation bandwidth and central frequency in our calculations were matched to those of the real data.
The chosen grid of RM values was $\langle{\rm RM}^{\rm mc}\rangle=\{0, 5, 10, 20, 50, 100, 200, 300, 500, 800, 1000\}$ rad m$^{-2}$.
In each Monte Carlo data set, the fake PAs were the products of randomisation of the calculated PA (from Eq.~\ref{eq:rmeq2})
within the window [$-\sigma_{\rm PA}^{\rm mc},\sigma_{\rm PA}^{\rm mc}$]. We ranged the value of $\sigma_{\rm PA}^{\rm mc}$ 
between 1 and 63 degrees, in 1-deg steps. For each combination of $\sigma_{\rm PA}^{\rm mc}$ and $\langle{\rm RM}^{\rm mc}\rangle$ values, 
we generated 2,000 Monte Carlo data sets.
Each of the generated MC data sets were then fitted using the fitting routine that was used with the real data.
The resulting RM values from the fits, ${\rm RM}^{\rm mc}$, were Gaussian distibuted around the value $\langle{\rm RM}^{\rm mc}\rangle$
that was used to generate that data set. For each distribution, we kept the RMS spread 
as the $\sigma_{\rm stat}$ corresponding to the particular pair of $\{\langle{\rm RM}^{\rm mc}\rangle,\sigma_{\rm PA}^{\rm mc}\}$ 
values that was used to generate the distribution.

The look-up table that was generated in this way allowed us to assign a statistical error to the ${\rm RM}^{\rm data}$ values 
of the real data, given the typical $\sigma_{\rm PA}^{\rm data}$ of the PAs, by matching 
--- as closely as possible ---  the values of $\sigma_{\rm PA}^{\rm data}$ and 
${\rm RM}^{\rm data}$ to those of $\sigma_{\rm PA}^{\rm mc}$ and $\langle{\rm RM}^{\rm mc}\rangle$. 
For values lying between the tabulated ones, we used a
two-dimensional cubic-spline interpolation of the table values.

\subsection{Systematic Error Determination} 
We made an effort to try and determine the systematic effects on the
data, which can be caused e.g.~by imperfect polarisation calibration
or RFI mitigation. The procedure followed was that above, but at the start of each fitting
process we averaged the number of frequency channels by a factor $K_f$, whose value we varied
before each fit. More specifically, three different factors were used, 32, 64
and 128, which guaranteed a minimum of 8 frequency channels in the fit
--- thus avoiding low statistics. It is expected that increasing the value of
$K_f$ will lead to increased s/n per frequency
channel; but it also means that there are fewer degrees of freedom in
the fit. Thus, the resulting value of the RM should stay the same,
unless there are systematic effects that alter each result. The
magnitude of those systematic effects can be estimated as the
bias-corrected standard deviation of those three values. Prior to this
estimate, each of the values was weighted by the inverse square of the
goodness-of-fit probability, $p_{\rm fit}$, arising from the Bayesian fitting. Hence,
we calculated the systematic error as
\begin{equation} 
\label{eq:syserr} 
\sigma_{\rm sys}=\sqrt{\frac{1}{2}\sum_{K_f=32}^{128}\left({\rm RM}_{K_f}-\left<{\rm RM}_{K_f}\right>\right)^2} 
\end{equation}
where $\left<{\rm RM}_{K_f}\right>$ is the weighted mean
with weights equal to $w_{K_f}=1/p^2_{\rm fit}$.

\vspace*{0.5cm}

\section{Results}
Of the 239 pulsars in our sample, our analysis produced RM values
for 150. The remaining 89 pulsars had either low s/n generally or
very little linear polarization. Of the 150 RMs obtained,
46 are values measured for the first time and 104 revised from their
previous values. The sample of previously unpublished RMs covers
over 200$^\circ$ Galactic longitude and is mostly confined within a 4 kpc
radius from the Sun. Apart from being a valuable addition to previous RM
databases, the new sample enriches our knowledge of RMs in the first
quadrant in particular ($0^\circ\leq\ell\lesssim 70^\circ$), which was
previously poorly sampled.
All resulting RM values and the respective combined errors
($\sigma_{{\rm RM}}=({\sigma_{\rm stat}^2+\sigma_{\rm sys}^2})^{1/2}$) are listed in
Table \ref{tab:RMresults}.

\setcounter{table}{0}
\begin{table*}
\caption{\label{tab:RMresults} The results from the herein analysis
(column 2), including the combined statistical and systematic errors
(column 3). The table is sorted by the RA of each pulsar. 
For compatibility with Fig.~\ref{fig:RMcomparisonPlot}, the ascending order according to 
$\ell$, of pulsars with previously published RM values, is given next to each pulsar's J2000 name (Column 1). 
Columns 4 and 5 show the RM values and their error that
were previously published for these pulsars; the relevant publications
are letter-coded, with the corresponding reference listed below the
table. The last two columns, column 6 and 7, list the average value of
the parallel component of the magnetic field, and its error, based on
Eq.~\ref{eq:avgBfield2}.}  
\centering 
\small
\begin{minipage}{200mm}
\begin{tabular}{@{}lrrrrrrrrrrr} 
\hline 
PSR                 &  $\ell$           &    $b$          &      DM         &         $d$       &       RM$_{\rm pub}$       &  $\sigma_{\rm RM}$   & RM$_{\rm new}$ & $\sigma_{\rm RM}$ & $B_{\parallel}$ & $\sigma_{B_{\parallel}}$\\ 
                    &  [deg]            &    [deg]        & [pc cm$^{-3}$]  &        [kpc]      &       [rad m$^{-2}$]       &  [rad m$^{-2}$]      & [rad m$^{-2}$] & [rad m$^{-2}$]    &     [$\mu$G]    &        [$\mu$G]         \\ 
\hline 

   J0034$-$0721$^{(18)}$     &       110.42     &     $-$69.82     &        11.38     &         0.41     &     $       9.8^{(a)}$     &          0.2     &         12.8     &            4     &       1.08     &       0.39    \\
   J0108$-$1431$^{(19)}$     &       140.93     &     $-$76.81     &         2.38     &          0.2     &     $      -0.3^{(b)}$     &          0.1     &          3.0     &            2     &       1.55     &       4.26    \\
   J0134$-$2937$^{(31)}$     &       230.25     &     $-$80.25     &        21.81     &         0.56     &     $        13^{(c)}$     &            2     &         15.4     &            3     &       0.79     &       0.15    \\
   J0151$-$0635$^{(20)}$     &       160.37     &        $-$65     &        25.66     &         1.22     &     $         2^{(a)}$     &            4     &          8.1     &            7     &       0.53     &       0.34    \\
   J0152$-$1637$^{(22)}$     &       179.31     &     $-$72.46     &        11.92     &         0.51     &     $         2^{(d)}$     &            1     &       $-$0.2     &            3     &       0.21     &       0.28    \\
   J0206$-$4028$^{(36)}$     &       258.60     &     $-$69.63     &         12.9     &         0.58     &     $        -4^{(d)}$     &            5     &          8.8     &           18     &       0.29     &       1.69    \\
   J0211$-$8159$^{(60)}$     &       299.59     &     $-$34.61     &        24.36     &            1     &     $        54^{(b)}$     &            9     &         30.0     &           16     &       1.77     &        0.8    \\
   J0255$-$5304$^{(39)}$     &       269.86     &     $-$55.31     &         15.9     &         0.73     &     $       -35^{(b)}$     &            3     &         32.1     &            3     &       2.48     &       0.27    \\
     J0304+1932$^{(21)}$     &       161.14     &     $-$33.27     &        15.74     &         0.62     &     $      -8.3^{(e)}$     &          0.3     &       $-$7.0     &            1     &    $-$0.55     &       0.11    \\
   J0401$-$7608$^{(54)}$     &       290.31     &     $-$35.91     &         21.6     &         0.68     &     $        19^{(d)}$     &         0.50     &         25.9     &            5     &       1.48     &       0.29    \\
   J0448$-$2749$^{(29)}$     &       228.35     &     $-$37.92     &        26.22     &         1.29     &     $        24^{(b)}$     &           17     &      $-$10.9     &           27     &       1.17     &       1.27    \\
   J0450$-$1248$^{(26)}$     &       211.08     &     $-$32.63     &        37.04     &         1.89     &     $        13^{(a)}$     &            5     &         15.5     &           18     &       1.23     &       0.59    \\
   J0459$-$0210$^{(25)}$     &       201.44     &     $-$25.68     &        21.02     &         0.92     &     $        18^{(c)}$     &            9     &         26.9     &           10     &       1.88     &       0.56    \\
   J0520$-$2553$^{(30)}$     &       228.43     &     $-$30.54     &        33.77     &         1.74     &     $        19^{(c)}$     &           15     &      $-$12.4     &           10     &    $-$0.22     &       0.36    \\
     J0525+1115$^{(23)}$     &       192.70     &     $-$13.25     &        79.34     &         3.11     &     $        37^{(f)}$     &            2     &         15.3     &           10     &       0.14     &       0.15    \\
   J0536$-$7543$^{(51)}$     &       287.16     &     $-$30.82     &         17.5     &         0.79     &     $        28^{(c)}$     &            2     &         25.2     &            1     &       1.76     &       0.07    \\
   J0540$-$7125$^{(44)}$     &       282.15     &     $-$31.24     &        29.41     &         1.27     &     $        43^{(c)}$     &           15     &         68.7     &           25     &       2.64     &       1.05    \\
   J0601$-$0527$^{(27)}$     &       212.20     &     $-$13.48     &        80.54     &         3.93     &     $      62.4^{(f)}$     &            2     &         64.0     &            2     &       0.99     &       0.03    \\
   J0624$-$0424$^{(28)}$     &       213.79     &      $-$8.04     &        70.83     &         2.78     &     $        42^{(a)}$     &            7     &         48.8     &            3     &       0.83     &       0.05    \\
     J0631+1036$^{(24)}$     &       201.22     &         0.45     &        125.4     &         3.67     &     $       137^{(b)}$     &            8     &        137.5     &            4     &       1.38     &       0.03    \\
   J0656$-$2228              &       233.66     &      $-$8.98     &        32.39     &          1.9     &     $               -$     &          $-$     &         38.3     &           12     &        1.9     &       0.45    \\
   J0719$-$2545              &       238.93     &      $-$5.83     &       253.89     &        45.32     &     $               -$     &          $-$     &        163.6     &           18     &       0.82     &       0.09    \\
   J0729$-$1836$^{(32)}$     &       233.76     &      $-$0.34     &        61.29     &          2.9     &     $        53^{(a)}$     &            6     &         51.5     &            4     &       1.01     &       0.09    \\
   J0738$-$4042$^{(34)}$     &       254.19     &      $-$9.19     &        160.8     &         2.64     &     $      14.5^{(g)}$     &          0.7     &         12.1     &          0.6     &        0.1     &    $<$0.01    \\
   J0742$-$2822$^{(33)}$     &       243.77     &      $-$2.44     &        73.78     &         2.07     &     $    149.95^{(c)}$     &         0.05     &        148.5     &          0.6     &       2.47     &       0.01    \\
   J0745$-$5353$^{(38)}$     &       266.63     &     $-$14.27     &        122.3     &         0.25     &     $       -72^{(d)}$     &            5     &      $-$71.0     &            4     &    $-$0.73     &       0.04    \\
   J0749$-$4247$^{(35)}$     &       257.07     &      $-$8.35     &       104.59     &         0.25     &     $        80^{(b)}$     &           30     &        124.8     &            5     &       1.51     &       0.06    \\
   J0835$-$4510$^{(37)}$     &       263.55     &      $-$2.79     &        67.99     &         0.29     &     $     31.38^{(h)}$     &         0.01     &         30.4     &          0.6     &       0.54     &       0.01    \\
   J0838$-$2621              &       248.81     &         8.98     &        116.9     &         2.63     &     $               -$     &          $-$     &         86.1     &           13     &       0.97     &       0.14    \\
   J0843$-$5022              &       268.50     &      $-$4.90     &       178.47     &         0.26     &     $               -$     &          $-$     &        155.5     &           23     &       1.02     &       0.16    \\
   J0849$-$6322              &       279.43     &     $-$12.17     &        91.29     &         3.25     &     $               -$     &          $-$     &     $-$124.8     &           15     &    $-$1.86     &       0.21    \\
   J0905$-$5127$^{(40)}$     &       271.63     &      $-$2.85     &       196.43     &         3.29     &     $       292^{(b)}$     &            3     &        291.0     &            2     &       1.83     &       0.01    \\
   J0907$-$5157$^{(41)}$     &       272.15     &      $-$3.03     &       103.72     &         0.93     &     $       -24^{(d)}$     &            3     &      $-$23.3     &            1     &    $-$0.29     &       0.02    \\
   J0941$-$5244$^{(42)}$     &       276.45     &         0.09     &       157.94     &         3.14     &     $      -243^{(b)}$     &            4     &     $-$230.5     &           21     &       $-$2     &       0.16    \\
   J1012$-$5857$^{(45)}$     &       283.71     &      $-$2.14     &        383.9     &         7.93     &     $        74^{(b)}$     &            6     &         72.2     &            8     &       0.23     &       0.03    \\
   J1017$-$5621              &       282.73     &         0.34     &        439.1     &         8.98     &     $               -$     &          $-$     &        364.9     &            7     &       1.04     &       0.02    \\
   J1036$-$4926$^{(43)}$     &       281.52     &         7.73     &       136.53     &         3.98     &     $       -11^{(c)}$     &            6     &      $-$38.0     &           10     &    $-$0.34     &       0.09    \\
   J1038$-$5831$^{(48)}$     &       286.28     &      $-$0.02     &        72.74     &         1.91     &     $       -53^{(d)}$     &           20     &      $-$15.0     &           10     &    $-$0.22     &       0.18    \\
   J1046$-$5813$^{(50)}$     &       287.07     &         0.73     &        240.2     &         4.37     &     $       125^{(b)}$     &           10     &        132.7     &            7     &       0.69     &       0.04    \\
   J1047$-$3032              &       273.49     &        25.13     &        52.54     &         2.37     &     $               -$     &          $-$     &      $-$36.9     &           23     &    $-$1.13     &       0.54    \\
   J1047$-$6709$^{(55)}$     &       291.31     &      $-$7.13     &       116.16     &         2.88     &     $       -73^{(b)}$     &            3     &      $-$79.3     &            2     &    $-$0.86     &       0.02    \\
   J1057$-$5226$^{(47)}$     &       285.98     &         6.65     &         30.1     &         0.72     &     $      47.2^{(i)}$     &          0.8     &         44.0     &            2     &       1.84     &       0.07    \\
   J1110$-$5637$^{(53)}$     &       289.28     &         3.53     &       262.56     &         5.62     &     $       426^{(b)}$     &           11     &        418.9     &            3     &       1.96     &       0.01    \\

\hline 
\end{tabular} 
\begin{tabular}{l l} 
a. & \citet{hl87}, \\ 
b. & \citet{hml+06}, \\ 
c. & \citet{hmq99}, \\ 
d. & \citet{qmlg95}, \\ 
e. & \citet{man74}, \\ 
f. & \citet{jhv+05}, \\ 
g. & \citet{vdhm97}, \\ 
h. & \citet{hmm+77}, \\ 
i. & \citet{tml93}, \\ 
j. & \citet{cmh91}, \\ 
k. & \citet{rl94} \\ 
\end{tabular}
\end{minipage}
\end{table*}

\setcounter{table}{0}
\begin{table*}
\caption{Continued} 
\centering 
\small
\begin{minipage}{200mm}
\begin{tabular}{@{}lrrrrrrrrrrr} 
\hline 
PSR                 &  $\ell$           &    $b$          &      DM         &         $d$       &       RM$_{\rm pub}$       &  $\sigma_{\rm RM}$   & RM$_{\rm new}$ & $\sigma_{\rm RM}$ & $B_{\parallel}$ & $\sigma_{B_{\parallel}}$\\ 
                    &  [deg]            &    [deg]        & [pc cm$^{-3}$]  &        [kpc]      &       [rad m$^{-2}$]       &  [rad m$^{-2}$]      & [rad m$^{-2}$] & [rad m$^{-2}$]    &     [$\mu$G]    &        [$\mu$G]         \\ 

\hline

   J1112$-$6613$^{(56)}$     &       293.19     &      $-$5.23     &        249.3     &         6.48     &     $       -94^{(b)}$     &           18     &     $-$132.0     &            4     &    $-$0.65     &       0.02    \\
   J1123$-$4844$^{(52)}$     &       288.30     &        11.61     &        92.92     &         2.86     &     $        -7^{(c)}$     &            5     &      $-$13.7     &            6     &    $-$0.12     &       0.07    \\
   J1133$-$6250$^{(57)}$     &       294.21     &      $-$1.30     &        567.8     &         12.1     &     $       880^{(b)}$     &           24     &        848.3     &            6     &       1.84     &       0.01    \\
   J1137$-$6700$^{(59)}$     &       295.79     &      $-$5.17     &       228.04     &          5.7     &     $        -1^{(b)}$     &           13     &      $-$11.0     &           16     &       0.02     &       0.09    \\
   J1141$-$3107$^{(46)}$     &       285.75     &        29.39     &        30.77     &         1.21     &     $       -60^{(b)}$     &           30     &      $-$26.1     &            7     &    $-$1.08     &       0.28    \\
   J1141$-$3322$^{(49)}$     &       286.59     &        27.27     &        46.45     &         1.91     &     $       -33^{(b)}$     &           14     &      $-$36.4     &            5     &    $-$0.98     &       0.12    \\
   J1146$-$6030$^{(58)}$     &       294.98     &         1.34     &       111.68     &         2.35     &     $        10^{(b)}$     &           17     &       $-$4.9     &            4     &    $-$0.04     &       0.04    \\
   J1159$-$7910$^{(62)}$     &       300.41     &     $-$16.55     &        59.24     &         1.91     &     $       -11^{(b)}$     &            9     &       $-$6.0     &           15     &       0.21     &       0.32    \\
   J1225$-$6408$^{(61)}$     &       300.13     &      $-$1.41     &        415.1     &        10.46     &     $       356^{(b)}$     &           23     &        336.6     &            4     &       0.99     &       0.01    \\
   J1231$-$4609              &       299.38     &        16.57     &           76     &         2.51     &     $               -$     &          $-$     &      $-$19.9     &            6     &    $-$0.31     &       0.09    \\
   J1236$-$5033              &       300.58     &        12.25     &       105.02     &         3.12     &     $               -$     &          $-$     &         49.2     &           13     &       0.48     &       0.16    \\
   J1240$-$4124              &       300.69     &        21.41     &         44.1     &         1.52     &     $               -$     &          $-$     &         15.1     &           13     &       0.75     &       0.36    \\
   J1253$-$5820$^{(63)}$     &       303.20     &         4.53     &       100.58     &         2.16     &     $        31^{(c)}$     &            5     &         17.7     &            1     &       0.21     &       0.01    \\
   J1305$-$6455$^{(64)}$     &       304.41     &      $-$2.09     &          505     &        12.13     &     $      -420^{(j)}$     &           10     &     $-$420.0     &            5     &    $-$1.03     &       0.01    \\
   J1306$-$6617$^{(65)}$     &       304.46     &      $-$3.46     &        436.9     &        12.38     &     $       387^{(b)}$     &           10     &        395.6     &            4     &       1.11     &       0.01    \\
   J1319$-$6056$^{(66)}$     &       306.31     &         1.74     &       400.94     &         7.85     &     $      -238^{(d)}$     &           25     &     $-$280.6     &            2     &    $-$0.86     &       0.01    \\
   J1320$-$3512              &       309.54     &        27.29     &        16.42     &         0.68     &     $               -$     &          $-$     &       $-$7.8     &            2     &    $-$0.53     &       0.13    \\
   J1320$-$5359$^{(68)}$     &       307.30     &         8.64     &         97.6     &         2.34     &     $       160^{(j)}$     &           10     &        141.1     &            4     &       1.78     &       0.05    \\
   J1326$-$5859$^{(69)}$     &       307.50     &         3.56     &        287.3     &         6.42     &     $      -580^{(j)}$     &           10     &     $-$579.6     &          0.9     &    $-$2.48     &    $<$0.01    \\
   J1327$-$6301$^{(67)}$     &       306.97     &      $-$0.43     &       294.91     &         5.26     &     $        96^{(b)}$     &           12     &         87.0     &            2     &       0.36     &       0.01    \\
   J1333$-$4449              &       310.77     &        17.40     &         44.3     &         1.38     &     $               -$     &          $-$     &      $-$71.5     &           17     &    $-$1.45     &       0.48    \\
   J1338$-$6204$^{(70)}$     &       308.37     &         0.31     &        640.3     &         9.81     &     $      -452^{(d)}$     &            8     &     $-$459.3     &            4     &    $-$0.89     &       0.01    \\
   J1339$-$4712              &       311.42     &        14.87     &         39.9     &         1.22     &     $               -$     &          $-$     &         17.0     &           11     &       0.56     &       0.34    \\
   J1340$-$6456              &       308.05     &      $-$2.56     &        76.99     &          1.7     &     $               -$     &          $-$     &      $-$37.1     &           23     &    $-$0.29     &       0.37    \\
   J1352$-$6803              &       308.61     &      $-$5.87     &        214.6     &         5.42     &     $               -$     &          $-$     &         30.0     &            7     &       0.14     &       0.04    \\
   J1356$-$5521$^{(72)}$     &       312.20     &         6.34     &       174.17     &         4.19     &     $       150^{(b)}$     &           22     &        101.1     &            4     &       0.71     &       0.03    \\
   J1403$-$7646              &       307.10     &     $-$14.49     &        100.6     &         3.26     &     $               -$     &          $-$     &         94.6     &           16     &       1.02     &       0.19    \\
   J1410$-$7404              &       308.35     &     $-$12.04     &        54.24     &         1.53     &     $               -$     &          $-$     &       $-$3.6     &            4     &    $-$0.05     &        0.1    \\
   J1413$-$6307$^{(71)}$     &       312.05     &      $-$1.72     &       121.98     &         2.34     &     $        45^{(b)}$     &            9     &         43.8     &            4     &       0.42     &       0.04    \\
   J1507$-$4352$^{(78)}$     &       327.34     &        12.46     &         48.7     &         1.34     &     $       -33^{(b)}$     &            4     &      $-$34.0     &            7     &    $-$0.73     &       0.18    \\
   J1507$-$6640              &       315.86     &      $-$7.30     &        129.8     &         3.56     &     $               -$     &          $-$     &      $-$40.0     &           13     &     $-$0.3     &       0.12    \\
   J1512$-$5759$^{(73)}$     &       320.77     &      $-$0.11     &        628.7     &         7.35     &     $       513^{(b)}$     &           16     &        510.0     &            7     &          1     &       0.01    \\
   J1514$-$4834              &       325.87     &         7.84     &         51.5     &         1.31     &     $               -$     &          $-$     &         18.1     &           14     &       0.81     &       0.34    \\
   J1517$-$4356              &       328.85     &        11.46     &        87.78     &         2.23     &     $               -$     &          $-$     &          1.0     &           19     &    $-$0.22     &       0.27    \\
   J1522$-$5829$^{(74)}$     &       321.63     &      $-$1.22     &        199.9     &         3.43     &     $       -18^{(d)}$     &           10     &      $-$24.2     &            2     &    $-$0.15     &       0.01    \\
   J1528$-$4109              &       332.11     &        12.67     &         89.5     &         2.63     &     $               -$     &          $-$     &         25.1     &            8     &       0.33     &       0.11    \\
   J1534$-$5405              &       325.46     &         1.48     &       190.82     &         3.37     &     $               -$     &          $-$     &      $-$69.5     &           12     &    $-$0.36     &       0.08    \\
   J1535$-$4114              &       333.18     &        11.82     &        66.28     &         1.95     &     $               -$     &          $-$     &         17.7     &            2     &       0.32     &       0.04    \\
   J1536$-$3602              &       336.55     &        15.84     &           96     &         3.16     &     $               -$     &          $-$     &      $-$23.8     &            9     &     $-$0.4     &       0.12    \\
   J1539$-$5626$^{(75)}$     &       324.62     &      $-$0.81     &       175.88     &         3.13     &     $       -16^{(d)}$     &            7     &      $-$18.0     &            2     &    $-$0.13     &       0.01    \\
   J1557$-$4258$^{(83)}$     &       335.27     &         7.95     &        144.5     &         4.59     &     $       -37^{(c)}$     &            2     &      $-$41.9     &            2     &    $-$0.35     &       0.02    \\
   J1600$-$5751$^{(76)}$     &       325.97     &      $-$3.70     &       176.55     &         3.49     &     $      -131^{(b)}$     &            8     &     $-$117.9     &           18     &    $-$0.96     &       0.12    \\
   J1603$-$5657$^{(77)}$     &       326.88     &      $-$3.31     &       264.07     &         5.14     &     $        27^{(c)}$     &            5     &         27.7     &            3     &       0.12     &       0.01    \\
   J1605$-$5257$^{(79)}$     &       329.73     &      $-$0.48     &         35.1     &         1.21     &     $         9^{(d)}$     &            3     &          1.0     &            2     &       0.04     &       0.06    \\
   J1611$-$5209$^{(80)}$     &       330.92     &      $-$0.48     &       127.57     &         4.35     &     $       -72^{(b)}$     &            6     &      $-$78.3     &            5     &    $-$0.74     &       0.05    \\
   J1614$-$3937              &          340     &         8.21     &       152.44     &         3.79     &     $               -$     &          $-$     &        133.0     &           16     &       1.13     &       0.13    \\
   J1614$-$5048$^{(81)}$     &       332.21     &         0.17     &        582.8     &         7.94     &     $      -451^{(b)}$     &            2     &     $-$452.7     &            5     &    $-$0.95     &       0.01    \\
   J1615$-$5537              &       329.04     &      $-$3.46     &       124.48     &         2.41     &     $               -$     &          $-$     &      $-$53.8     &           16     &    $-$0.55     &       0.16    \\
   J1630$-$4733$^{(84)}$     &       336.40     &         0.56     &          498     &         5.65     &     $      -338^{(b)}$     &            8     &     $-$348.2     &            6     &    $-$0.86     &       0.01    \\
   J1633$-$4453$^{(88)}$     &       338.73     &         1.98     &        474.1     &         7.12     &     $       139^{(b)}$     &           17     &        159.0     &            6     &        0.4     &       0.01    \\
   J1633$-$5015$^{(82)}$     &       334.70     &      $-$1.57     &       398.41     &         5.68     &     $       307^{(d)}$     &           12     &        406.1     &            2     &       1.26     &       0.01    \\
   J1637$-$4553$^{(86)}$     &       338.48     &         0.76     &       193.23     &         3.16     &     $        12^{(b)}$     &            4     &         10.2     &            5     &       0.05     &       0.03    \\
   J1639$-$4604$^{(87)}$     &       338.50     &         0.46     &       258.91     &         3.76     &     $       -60^{(b)}$     &           30     &      $-$28.3     &           12     &    $-$0.13     &       0.06    \\
   J1640$-$4715$^{(85)}$     &       337.71     &      $-$0.44     &        591.7     &         6.48     &     $      -398^{(b)}$     &           22     &     $-$411.3     &           12     &    $-$0.86     &       0.02    \\
   J1641$-$2347              &       355.83     &        14.71     &         27.7     &         1.01     &     $               -$     &          $-$     &      $-$22.2     &            4     &    $-$1.02     &       0.17    \\

\hline 
\end{tabular} 
\end{minipage}
\end{table*}

\setcounter{table}{0}
\begin{table*}
\caption{Continued} 
\centering 
\small
\begin{minipage}{200mm}
\begin{tabular}{@{}lrrrrrrrrrrr} 
\hline 
PSR                 &  $\ell$           &    $b$          &      DM         &         $d$       &       RM$_{\rm pub}$       &  $\sigma_{\rm RM}$   & RM$_{\rm new}$ & $\sigma_{\rm RM}$ & $B_{\parallel}$ & $\sigma_{B_{\parallel}}$\\ 
                    &  [deg]            &    [deg]        & [pc cm$^{-3}$]  &        [kpc]      &       [rad m$^{-2}$]       &  [rad m$^{-2}$]      & [rad m$^{-2}$] & [rad m$^{-2}$]    &     [$\mu$G]    &        [$\mu$G]         \\ 
 
\hline

   J1646$-$4346$^{(90)}$     &       341.11     &         0.97     &        490.4     &         5.79     &     $       -62^{(b)}$     &            7     &      $-$24.2     &           18     &    $-$0.11     &       0.04    \\
   J1650$-$1654              &         2.86     &        17.23     &        43.25     &         1.47     &     $               -$     &          $-$     &          7.2     &           14     &       0.54     &       0.41    \\
   J1651$-$7642              &       315.15     &     $-$19.95     &           80     &         2.99     &     $               -$     &          $-$     &         54.1     &            6     &        0.8     &        0.1    \\
   J1652$-$1400              &         5.60     &        18.58     &         49.5     &         1.71     &     $               -$     &          $-$     &      $-$49.0     &           10     &    $-$1.24     &       0.24    \\
   J1653$-$3838$^{(93)}$     &       345.88     &         3.27     &        207.2     &         3.66     &     $       -74^{(b)}$     &            6     &      $-$81.7     &            3     &    $-$0.47     &       0.02    \\
   J1655$-$3048              &       352.24     &         7.88     &        154.3     &         3.82     &     $               -$     &          $-$     &      $-$63.1     &           10     &    $-$0.51     &       0.08    \\
   J1700$-$3312$^{(97)}$     &       351.06     &         5.49     &       166.97     &         3.62     &     $       -15^{(c)}$     &            3     &      $-$25.3     &            4     &    $-$0.18     &       0.03    \\
   J1701$-$3726$^{(95)}$     &       347.76     &         2.83     &        303.4     &         5.17     &     $      -602^{(b)}$     &            8     &     $-$605.9     &            2     &    $-$2.46     &       0.01    \\
   J1701$-$4533$^{(91)}$     &       341.36     &      $-$2.18     &          526     &         9.71     &     $        17^{(b)}$     &           13     &          4.0     &            4     &       0.01     &       0.01    \\
   J1703$-$4851$^{(89)}$     &       338.99     &      $-$4.51     &       150.29     &         2.99     &     $        -4^{(b)}$     &           24     &      $-$46.2     &            5     &    $-$0.36     &       0.04    \\
   J1705$-$1906$^{(3)}$      &         3.19     &        13.03     &        22.91     &         0.88     &     $        -9^{(a)}$     &            4     &      $-$19.2     &            1     &    $-$1.08     &       0.06    \\
   J1707$-$4053$^{(92)}$     &       345.72     &      $-$0.20     &          360     &         4.49     &     $      -207^{(d)}$     &           25     &        168.2     &            4     &       0.56     &       0.01    \\
   J1717$-$5800              &       332.67     &     $-$11.48     &       125.22     &         3.47     &     $               -$     &          $-$     &         39.0     &           20     &        0.5     &        0.2    \\
   J1719$-$4006$^{(94)}$     &       347.65     &      $-$1.53     &        386.6     &         5.13     &     $      -234^{(b)}$     &           31     &     $-$217.9     &           17     &    $-$0.73     &       0.05    \\
   J1721$-$3532$^{(99)}$     &       351.69     &         0.67     &          496     &         5.62     &     $       205^{(d)}$     &           75     &        158.9     &            4     &        0.4     &       0.01    \\
   J1722$-$3632$^{(96)}$     &       350.93     &            0     &        416.2     &         4.35     &     $      -307^{(b)}$     &            8     &     $-$332.8     &            9     &    $-$0.96     &       0.03    \\
   J1733$-$2228$^{(6)}$      &         4.03     &         5.75     &        41.14     &         1.17     &     $        -9^{(a)}$     &            4     &      $-$12.0     &            3     &    $-$0.33     &        0.1    \\
   J1733$-$3716$^{(98)}$     &       351.58     &      $-$2.28     &        153.5     &          2.8     &     $      -330^{(b)}$     &            6     &     $-$335.0     &            2     &    $-$2.68     &       0.02    \\
   J1737$-$3555              &       353.17     &      $-$2.27     &        89.41     &         1.76     &     $               -$     &          $-$     &         50.0     &            4     &        0.7     &       0.06    \\
   J1739$-$1313              &        12.79     &         9.30     &         58.2     &         1.61     &     $               -$     &          $-$     &         34.1     &            2     &       0.74     &       0.05    \\
   J1740$-$3015$^{(103)}$    &       358.29     &         0.24     &       152.15     &         2.73     &     $      -157^{(k)}$     &            2     &     $-$168.0     &          0.7     &    $-$1.36     &       0.01    \\
   J1742$-$4616              &       344.79     &      $-$8.46     &       115.96     &         2.69     &     $               -$     &          $-$     &      $-$37.9     &            6     &    $-$0.39     &       0.07    \\
   J1743$-$3150$^{(101)}$    &       357.30     &      $-$1.15     &       193.05     &         3.31     &     $      -240^{(b)}$     &           12     &     $-$215.0     &            4     &    $-$1.37     &       0.02    \\
   J1743$-$4212              &       348.38     &      $-$6.46     &       131.94     &         2.87     &     $               -$     &          $-$     &      $-$20.5     &            6     &    $-$0.16     &       0.05    \\
   J1749$-$3002$^{(104)}$    &       359.46     &      $-$1.24     &        509.4     &         5.83     &     $      -313^{(d)}$     &            7     &     $-$289.7     &            3     &     $-$0.7     &       0.01    \\
   J1750$-$3157$^{(102)}$    &       357.98     &      $-$2.52     &       206.34     &         3.82     &     $       111^{(b)}$     &            8     &        108.6     &           14     &       0.72     &       0.08    \\
   J1757$-$2421$^{(7)}$      &         5.28     &         0.05     &       179.45     &          4.4     &     $        -9^{(b)}$     &            9     &         15.9     &            5     &        0.1     &       0.03    \\
   J1801$-$2920$^{(1)}$      &         1.44     &      $-$3.25     &       125.61     &         2.77     &     $       -60^{(d)}$     &           10     &      $-$61.8     &            3     &     $-$0.6     &       0.03    \\
   J1803$-$2712$^{(4)}$      &         3.49     &      $-$2.53     &        165.5     &         3.38     &     $      -147^{(d)}$     &           18     &     $-$165.0     &            6     &    $-$1.22     &       0.04    \\
   J1805$-$0619              &        21.99     &         7.22     &       146.22     &         3.98     &     $               -$     &          $-$     &         82.4     &           14     &       0.69     &       0.12    \\
   J1808$-$0813$^{(12)}$     &        20.63     &         5.75     &       151.27     &         3.76     &     $        73^{(c)}$     &            7     &         76.8     &            5     &       0.63     &       0.04    \\
   J1808$-$3249              &       359.04     &      $-$6.11     &       147.37     &         3.64     &     $               -$     &          $-$     &        289.4     &            6     &       2.46     &       0.05    \\
   J1811$-$0154              &        26.61     &         8.03     &        148.1     &         4.44     &     $               -$     &          $-$     &         46.5     &           11     &       0.42     &       0.09    \\
   J1817$-$3837$^{(100)}$    &       354.68     &     $-$10.41     &       102.85     &         2.51     &     $        89^{(b)}$     &            6     &        102.9     &            2     &       1.23     &       0.02    \\
   J1820$-$1818$^{(11)}$     &        13.20     &      $-$1.72     &          436     &         7.04     &     $       -60^{(b)}$     &           24     &      $-$69.8     &           12     &    $-$0.18     &       0.03    \\
   J1822$-$2256$^{(8)}$      &         9.35     &      $-$4.37     &        121.2     &         2.99     &     $       142^{(d)}$     &            4     &        124.0     &            3     &       1.26     &       0.03    \\
   J1835$-$1106              &        21.22     &      $-$1.51     &       132.68     &         2.83     &     $               -$     &          $-$     &         42.3     &            3     &        0.4     &       0.02    \\
   J1837$-$0045              &        30.67     &         2.75     &        86.98     &         2.48     &     $               -$     &          $-$     &        130.3     &           17     &       1.98     &       0.23    \\
     J1837+1221              &        42.41     &         8.74     &        100.6     &         3.91     &     $               -$     &          $-$     &        172.9     &           24     &       2.13     &        0.3    \\
   J1837$-$1837              &        14.81     &      $-$5.50     &       100.74     &         2.73     &     $               -$     &          $-$     &        137.8     &            8     &       1.63     &        0.1    \\
   J1852$-$2610$^{(9)}$      &         9.46     &     $-$11.92     &        56.81     &         1.75     &     $       -21^{(c)}$     &            1     &      $-$21.2     &            4     &    $-$0.48     &       0.08    \\
   J1901$-$0906$^{(13)}$     &        25.98     &      $-$6.44     &        72.68     &         2.13     &     $        44^{(c)}$     &           12     &         29.0     &            2     &       0.51     &       0.03    \\
   J1901$-$1740              &        18.14     &     $-$10.07     &         24.4     &         0.95     &     $               -$     &          $-$     &         62.8     &           33     &       1.11     &       1.66    \\
     J1904+0004$^{(16)}$     &        34.45     &      $-$2.81     &       233.61     &         5.74     &     $       306^{(b)}$     &            9     &        289.0     &            6     &       1.53     &       0.03    \\
     J1919+0134              &        37.58     &      $-$5.56     &        191.9     &         6.17     &     $               -$     &          $-$     &         47.0     &            4     &        0.3     &       0.02    \\
   J1932$-$3655$^{(2)}$      &         2.07     &     $-$23.55     &        59.88     &         2.09     &     $         6^{(c)}$     &            3     &       $-$7.7     &            3     &    $-$0.14     &       0.07    \\
     J1943+0609              &        44.47     &      $-$8.64     &        70.76     &         3.02     &     $               -$     &          $-$     &      $-$11.1     &           15     &    $-$0.26     &       0.26    \\
   J2006$-$0807$^{(15)}$     &        34.10     &     $-$20.30     &        32.39     &         1.23     &     $       -52^{(a)}$     &            5     &      $-$61.6     &            3     &    $-$2.32     &       0.13    \\
   J2038$-$3816$^{(5)}$      &         3.85     &     $-$36.74     &        33.96     &         1.36     &     $        68^{(b)}$     &           18     &         38.4     &           14     &       1.49     &        0.5    \\
   J2048$-$1616$^{(14)}$     &        30.51     &     $-$33.08     &        11.46     &         0.56     &     $       -10^{(a)}$     &         0.07     &      $-$10.2     &          0.8     &    $-$1.18     &        0.1    \\
   J2108$-$3429$^{(10)}$     &         9.70     &     $-$42.16     &        30.22     &         1.21     &     $        50^{(c)}$     &           20     &         39.3     &           12     &       0.98     &       0.51    \\
   J2346$-$0609$^{(17)}$     &        83.80     &     $-$64.01     &         22.5     &         0.94     &     $        -5^{(c)}$     &            1     &       $-$4.1     &            6     &    $-$0.33     &        0.3    \\

\hline 
\end{tabular} 
\end{minipage}
\end{table*}

\begin{figure}
\vspace*{10pt} 
\includegraphics[width=0.47\textwidth]{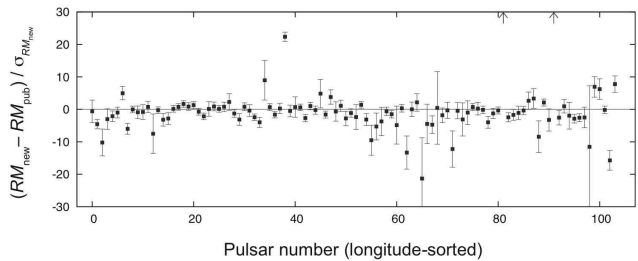} 
\caption{\label{fig:RMcomparisonPlot} Residuals' plot between
previously published RMs (RM$_{\rm pub}$) and those calculated from
our analysis (RM$_{\rm new}$), for 108 pulsars, sorted by 
increasing Galactic longitude ($\ell$). The order number for each 
pulsar is shown in parentheses, in column 1 of Table~\ref{tab:RMresults}. 
The RM residuals were calculated by subtracting the published from the new RM value
and dividing the result by the standard deviation of the new RM. The
two arrows at the top of the plot correspond to RM measurements for PSRs J1633$-$5015 (leftmost)
and J1707$-$4053 (rightmost), for which the residuals were outside the plot's range.}
\end{figure}

Fig.~\ref{fig:RMcomparisonPlot} shows a comparison between our sample of RMs and previous results from the 
literature. The results agree within the error bars for the majority of the sample.
Some objects show disagreement at the level of up to 10~rad m$^{-2}$.
This could simply be due to the measurement techniques used and/or an
incomplete estimate of the error bars. However, it is known that
systematic changes in RM at this sort of level can occur over periods
of years as the pulsar traverses through the interstellar medium
(e.g. in the Vela pulsar, Johnston et al. 2005\nocite{jhv+05}).
These differences would be worth following up in more detail.
A few pulsars have extremely discrepant RMs, the nature of which is
not entirely clear but may be as simple as typographical errors in the
original reports. In particular, the  RMs of PSRs J0255$-$5304, J1633$-$5015
and J1707$-$4053 are wildly at odds with literature values (see Fig.~\ref{fig:RMcomparisonPlot}).

\section{Magnetic Field Modelling}
The value of the rotation measure for each pulsar from observations can be
used to provide an estimate of the average magnitude of the component of the
regular Galactic magnetic field that is directed along the {\em los} to the
pulsar:
\begin{equation} 
\label{eq:avgBfield} 
\left<B_{\parallel}\right> = \frac{\int_{0}^{d}n_e(s)B_\parallel ds}{\int_{0}^{d}n_e(s)}
\end{equation} 
where the denominator is the column-density integral to the pulsar, called the dispersion measure, DM.
Hence, 
\begin{equation} 
\label{eq:avgBfield2} 
\left<B_{\parallel}\right>=1.232 \  \left(\frac{{\rm RM}}{{\rm rad} \ {\rm m}^{-2}}\right)\left(\frac{{\rm DM}}{{\rm pc} \ {\rm cm}^{-3}}\right)^{-1}
\end{equation} 
resulting in a value for $B_{\parallel}$ in $\mu$G.

The form of the Galactic free-electron density, $n_e(s)$, is derived
from modelling. In this work, we assumed the density distribution
described in the NE2001 model by \citet{cl01} --- other models include
e.g.~the TC93 model by \citet{tc93}. Furthermore, in what follows, we 
will use the Galactocentric Cartesian coordinate system of the NE2001 model, shown in Fig.~\ref{fig:losvectors}. 
All radii in our modelling correspond to Galactocentric distances, with the azimuth angle, 
$\theta$, measured clockwise from the Galactic centre--Sun direction.  

For the observed sample of pulsars, the value of
$\left<B_{\parallel}\right>$ is shown in column 6 of
Table~\ref{tab:RMresults}. Despite the fact that pulsar RMs allow us
to probe only the average magnitude of the GMF along the
{\em los}, the numerous, scattered pulsar positions across the
Galactic disc --- as well as a number of them in the Galactic halo ---
makes up for a powerful method to test the various theoretical models
that describe its shape and magnitude.

\subsection{Regular GMF models}
\label{subsec:RegFieldModels}
Below are four analytical models which we chose to test
for consistency against the data. The first three are described in
\citet{kst07} and have the common component of a logarithmic-spiral
disc field (first proposed by \nocite{sk80}Simard-Normandin \& Kronberg 1980). The last model is the
pure dipolar--toroidal field (produced by an A0 dynamo), which is
considered the most likely configuration for the magnetic field of the
halo (e.g.~\nocite{hmbb97}Han et al. 1997, \nocite{han02}Han 2002).
In the description of the spiral models, we have indicated in the
subsection title whether the model has a bisymmetric spiral (BSS) or 
antisymmetric spiral (ASS) configuration and whether the field
is antisymmetric (indicated by ``-A'') or symmetric
(indicated by ``-S'') with respect to the GP ($z=0$).

Previous efforts to check the consistency between the available models
and the data include \citet{hmq99}, who concluded that an
ASS with no reversals fits well pulsars
at high latitudes ($\mid b\mid>8^\circ$), whereas the disc field can
be better described using a bisymmetric-spiral configuration (BSS); a
concentric-ring model \citep{rk89} was the least supported by the
data. \citet{id99} have also tried to fit 131 pulsars to a
concentric-ring and a BSS model: the latter gave the best overall
fit. Nevertheless, it is worth mentioning that both models supported a
stronger field in the interarm regions compared to that in the arms
and that the general conclusion was that the field is reversed as one 
crosses the boundary between the arm and interarm regions.

\subsubsection{TT Model (BSS-A)}
The first model we explored was based on the model of \citet{tt02} (TT). 
TT developed their model primarily to explain the trajectories of
UHECRs from Active Galactic Nuclei (AGN) deflected by the GMF. Its shape
resembles the spiral shape of the distribution of the optical matter
in the Milky Way. The magnetic field in cylindrical coordinates is
\begin{equation} 
\begin{split}
\label{eq:TTfield} 
\boldsymbol{B} =
b(r)\cos\left[\theta-\frac{1}{\tan{p}}\ln\left(\frac{r}{r_0}\right)\right]&f(z)\times \\ 
\times &(\sin p \hat{\boldsymbol{e}}_r+\cos p \hat{\boldsymbol{e}}_\theta) \\
\end{split}
\end{equation} 
where $p$ is the {\em pitch angle}, which describes how tightly wound
the spiral arms are: the published value for this model is
$p=-8^\circ$. The radius $r_0$ represents the
Galactocentric distance along the line between the Sun and the Galactic centre (hereafter GC) to
the field maximum, $\mid\boldsymbol{B}(r,0)\mid$. TT express
$r_0$ in terms of the heliocentric distance to the closest magnetic-field reversal,
$d_0$: $r_0=(r_\odot+d_0)\exp(-\frac{1}{2}\pi\tan{p})$, where
$d_0=-0.5$ kpc. A negative $d_0$ implies that the closest reversal
occurs between the Sun and the GC ($\ell=0^\circ$), whereas positive
values of $d_0$ place the closest reversal towards the Galactic
anticentre ($\ell=180^\circ$). The
functions $b(r)$ and $f(z)$ are characteristic of such ``spiral''
magnetic-field models and respresent the radial and vertical profile
of the field's magnitude, respectively: for the TT model, $b(r>r_{\rm
core})=b(r_\odot)\times(r_\odot/r)$, where $b(r_\odot)=1.4$ $\mu$G,
and $b(r\leq r_{\rm core})={\rm const}$, where $r_{\rm core}=4$
kpc. The vertical decay is given by the exponential law $f(z)={\rm
sign}(z)\exp{(-\mid z\mid/z_0)}$, where it is typically assumed that the value for the scale height of the
Galactic halo, $z_0$, is 1.5 kpc.

\subsubsection{HMR Model (BSS-S)}
Another spiral model which we tested against our data was based on the model of \citet{hmr99} (HMR).
This model shares the same magnetic-field shape in the Galactic disc with that of the TT model but
differs in the field symmetry with respect to the GP, the latter being symmetric for the HMR model. 

HMR describe their spiral as somewhat less
tightly wound than that of TT, with $p=-10^\circ$, whereas the first
magnetic-field reversal happens at $d_0\approx-0.5023$ kpc. The radial
magnetic-field decay in this model is described by the
hyperbolic-tangent law $b(r)=(3r_\odot/r)\tanh^3(r/r_{\rm core})$,
where $r_{\rm core}=2$ kpc. The vertical suppression in the HMR model
follows the two-scale law $f(z)=1/[2\cosh(z/z_1)]+1/[2\cosh(z/z_2)]$,
where the scale heights $z_1$ and $z_2$ are typically set equal to 0.3 and 4
kpc, respectively.

\subsubsection{PS Model (BSS-S)}
The last spiral model that we used is a variation of the BSS-S model by \citet{ps03} (PS). The shape of the disc is
identical to that described in the TT model, the only differences
being the smaller halo size, with scale height $z_0=0.2$ kpc, and the
higher field magnitude, which we normalised to $b(r_\odot)=2$ $\mu$G.

In addition to the disc and halo components, PS considered a toroidal
contribution, $\boldsymbol{B}_T$, in the form of circular discs either side of the GP, 
with the vertical profile of the field following
a Lorentzian distribution:

\begin{equation} 
\label{eq:PStorfield} 
\begin{split}
\boldsymbol{B}_{T} &= -{\rm sign}(z)B_{T,{\rm max}}(r_\odot) \times \\
\times & \frac{H(r_\odot-r)+H(r-r_\odot){\rm
e}^{\frac{r_\odot-r}{r_\odot}}}{1+\left(\frac{\mid
z\mid-h_T}{w_T}\right)^2} (\cos\theta
\hat{\boldsymbol{x}}-\sin\theta\hat{\boldsymbol{y}})\\
\end{split}
\end{equation} 
where $B_{T,{\rm max}}(r_\odot)=1.5$ $\mu$G is the field maximum above
the GP; $h_T=1.5$ kpc is the height of the field maximum and $w_T=0.3$
kpc is the half-width of the Lorentzian field profile; finally, $H(x)$
is the Heaviside step function.

A final, dipolar component is added to the above two, with Cartesian
components
\begin{equation} 
\label{eq:PSdipolefield} 
\begin{split}
B_{x} &= -3\frac{\mu_{\rm G}}{R^3}\cos\phi \sin\phi \sin\theta
\hspace{5.5cm} \\ B_{y} &= -3\frac{\mu_{\rm G}}{R^3}\cos\phi \sin\phi
\cos\theta \\ B_{z} &= \frac{\mu_{\rm G}}{R^3}(1-3\cos^2\phi)
\end{split}
\end{equation} 
where $R=({r^2+z^2})^{1/2}$, $\theta=\tan^{-1}(x/y)$ and
$\phi=\cos^{-1}(z/R)$ are the Galactocentric spherical coordinates;
$\mu_{\rm G}=123$ $\mu$G kpc$^3$ is the magnetic dipole moment. The
singularity at $R=0$ is avoided by setting $B_z=-100$ $\mu$G for
$R<500$ pc.

\subsubsection{Dipolar--Toroidal Model}
In addition to the spiral models, we used a pure Dipolar--Toroidal magnetic
field (see e.g.~\nocite{hmbb97}Han et al. 1997). The total field, which is the
sum of the toroidal and the dipolar fields, is given in spherical
coordinates as
\begin{equation} 
\label{eq:ToroidalDipoleField} 
\begin{split}
\boldsymbol{B} = \boldsymbol{B}_{\rm dipole} &+ \boldsymbol{B}_{\rm
toroidal} = \\ =&\frac{m}{R^3}(2\cos\theta \hat{\boldsymbol{e}}_R +
\sin\theta\hat{\boldsymbol{e}}_\theta)+{\rm
sign}(z)\frac{n}{R}\hat{\boldsymbol{e}}_\theta
\end{split}
\end{equation} 
The parameters $m$ and $n$ represent the relative magnitude of the dipolar and toroidal
field components, respectively, at a given point in the Galaxy, weighted by the
Galactocentric distance of that point. Recent work has reported a
value of $m=245^{+53}_{-56}$ $\mu$G kpc$^3$ and $n=4.8^{+0.8}_{-0.7}$
$\mu$G kpc via $\chi^2$ fitting of pulsar RMs from a sample of
nearby pulsars with $d\lesssim 0.5$ kpc \citep{ath04}.

\medskip

Apart from the dipolar field of the PS
model, none of the above model fields includes a vertical
component. This is characteristic of most GMF models, although it is
believed that a dipolar structure at the GC may be the reason for a
weak $B_z\approx 0.2$ $\mu$G at $r_\odot$ \citep{hq94}.

\begin{figure*} 
\vspace*{10pt}
\includegraphics[width=1\textwidth]{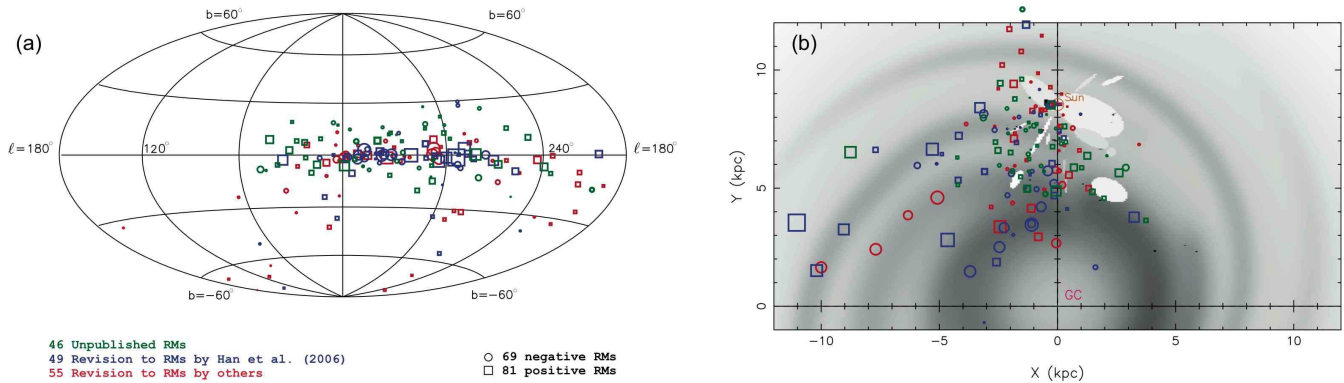}
\caption{\label{fig:HammerAndPlane} (a) Hammer-Aitoff (edge-on) and (b) Galactic-plane 
(face-on) projections of the observed sample of pulsars which were used in our 
analysis. The Galactic-plane projection also includes a gray-scale 
map of the free-electron density distribution according to the NE2001 model. The
associated symbols correspond to 150 good-quality RM measurements. Each
symbol's linear dimension is proportional to $({\mid {\rm RM}\mid})^{1/2}$ derived from
each measurement: positive RM measurements are represented with squares;
negative ones, with circles. The largest square corresponds to ${\rm RM}=848.3$ rad
m$^{-2}$ and the largest circle, to ${\rm RM}=-605.9$ rad m$^{-2}$. In addition,
the plot further classifies the total sample into RMs from pulsars without a
previous measurement (green), values from pulsars that had an RM recently
published by Han et al. (blue), and values from pulsars with previously
published RMs by other authors (red).}
\end{figure*}

\section{The Large-Scale Structure of the Galactic Magnetic Field Revisited}
Despite the conclusive evidence for the large-scale GMF direction
within $d\sim 1$ kpc (e.g.~the CW direction of the local field and the
field reversal in the first quadrant (Q1) of the Sagittarius-Carina
arm, \nocite{ls89}Lyne \& Smith 1989), past analyses of the available database of pulsar
RMs have resulted in contradicting opinions with respect to the
field's structure at larger distances: \citet{mwkj03} studied 45
pulsar RMs towards the Perseus arm and concluded that no magnetic
field reversal is evident to a 5--6 kpc distance; more recently, a
broader analysis of 223 southern pulsars by Han et al. concluded that
the arm regions maintain a counter-clockwise (CCW) magnetic field
whereas the data suggest
that the field is clockwise (CW) in the interarm regions
\citep{hml+06}. In addition to the usefulness of Galactic pulsars in studying the GMF,
a number of authors have included extragalactic sources in their analyses. In 2003, \citet{btwm03}
combined 262 RMs of extragalactic radio sources from the Canadian Galactic Plane Survey (CGPS; \nocite{tlj99}Taylor et al. 1999)
with 12 RMs of pulsars observed by \citet{mwkj03} and others, to conclude that there are no reversals beyond the solar radius.
Very recently, \citet{bhg+07} measured the RMs of 148
extragalactic radio sources from the Southern Galactic Plane Survey (SGPS; \nocite{mdg+05}McClure--Griffiths et al. 2005) and 
combined them with 120 known pulsar RMs, from \citet{hml+06}, \citet{hmq99} and \citet{tml00}, within the SGPS region ($253^\circ \leq \ell \leq 358^\circ$). They
concluded that no reversals exist beyond the radius of the
Sagittarius--Carina arm, in the region surveyed by the SGPS, and that the field maintains a CW
direction in that region. However, they found clear evidence of a
reversal in the interarm region between the Sagittarius--Carina and
Scutum--Crux arms, where the field is CCW.

In order to shed more light on the problem of field reversals beyond
the solar neigbourhood, we analysed the sample of 150
pulsar RMs shown in Table \ref{tab:RMresults}. Amongst them, there are 57
previously unpublished values that correspond to 52 pulsars, of which
more than 90\% are distributed within $d=4$ kpc, but with a longitudinal
spread covering more that $200^\circ$ across the sky. The longitudinal and
latitudinal pulsar distribution in the sky is graphically presented in
Fig.~\ref{fig:HammerAndPlane}a, which shows a Hammer-Aitoff projection of
all pulsar positions and their relative RM magnitude. Furthermore, a
two-dimensional projection of the GP in Fig.~\ref{fig:HammerAndPlane}b shows
the radial distribution of the pulsars.

A straightforward method to assess the radial profile of $\left<
B_{\parallel}\right>$ is to plot its value against distance, along
different lines-of-sight. According to \citet{ls89}, the magnetic
field in the region between distance $d_1$ and $d_2$ along the
{\em los} can be assessed from the ratio of the RM and DM
gradients over that region: i.e.
\begin{equation} 
\label{eq:RMDM_gradients} 
\left<B_{\parallel}\right>_{d1-d2} = 1.232\frac{\Delta {\rm RM}}{\Delta {\rm DM}}
\end{equation} 

\begin{figure} 
\vspace*{10pt} \includegraphics[width=0.47\textwidth]{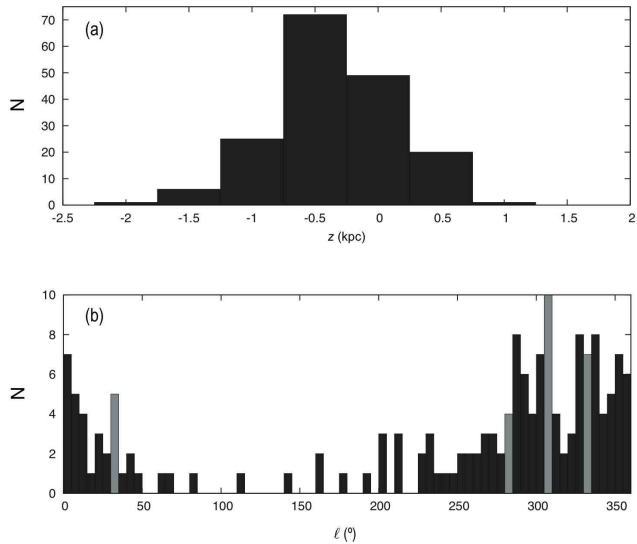}
\caption{\label{fig:zldistrib} The distributions of (a) the $z$-height
and (b) the Galactic longitude for the pulsar sample used in the
herein analysis. The light-gray longitude bins correspond to the four
$5^\circ$ sectors that were checked for field reversals (see section
\ref{subsec:locOfRevs}).}
\end{figure}

Ideally, one would need an infinite number of pulsars in order to
represent the actual fluctuations of $\left< B_{\parallel}\right>_d$
with distance, $d$. In practice, this method is limited by the number
of pulsars lying within a solid angle of a certain longitude and
latitude width, $\Delta \ell$ and $\Delta b$, respectively. However,
since most pulsars in our sample lie within $\mid z\mid=2$
kpc of the GP (see Fig.~\ref{fig:zldistrib}a), and because, according
to the $B$-field models described above, the disc field does not
extend beyond that $z$-height,
it was more appropriate to plot the $\left< B_{\parallel}\right>_d$
function inside the zone defined by $\mid z\mid\leq 2$ kpc, instead of
using a latitude constraint.

In principle, such plots can be drawn for a number of longitude
sectors across the sky. One convenient way, however, to map the
locations of the field reversals in a single plot is an image
map of $\left< B_{\parallel}\right>_d$, which is constructed by
projecting the pulsar locations onto the Galacting plane and binning
the average field values, along with the pulsar coordinates, in a
polar grid (referenced at the Sun). A linear interpolation of the
available field values is also introduced to smooth the GMF
gradients between bins. This projection introduces an uncertainty in
the $z$ direction, which also propagates into the calculation of
$\left< B_{\parallel}\right>$ through the scale factor
$\gamma=1/({1+(z/\rho)^2})^{1/2}$, where $\rho=({d^2-z^2})^{1/2}$ is the
polar radius. So, for a planar GMF, $\boldsymbol{B}(x,y)$, the
projected value of the average field expressed by Eq.~\ref{eq:avgBfield}
is modified as follows as one moves away from the GP:
\begin{equation} 
\label{eq:Bll_scaling} 
\left<B_{\parallel}\right> = \frac{\int_0^{\gamma d}n_e(s)\gamma
B_{\parallel}(s)ds}{\int_0^{\gamma d}n_e(s)ds}
\end{equation} 
where it is also assumed that the field direction does not change
between the $z$-boundaries. Clearly, the projected field values are
representative of the actual ones as long as $z$ is chosen small
enough for $\gamma\approx 1$. 

\subsection{Location of Reversals}
\label{subsec:locOfRevs}
\begin{figure*} 
\includegraphics[width=0.65\textwidth]{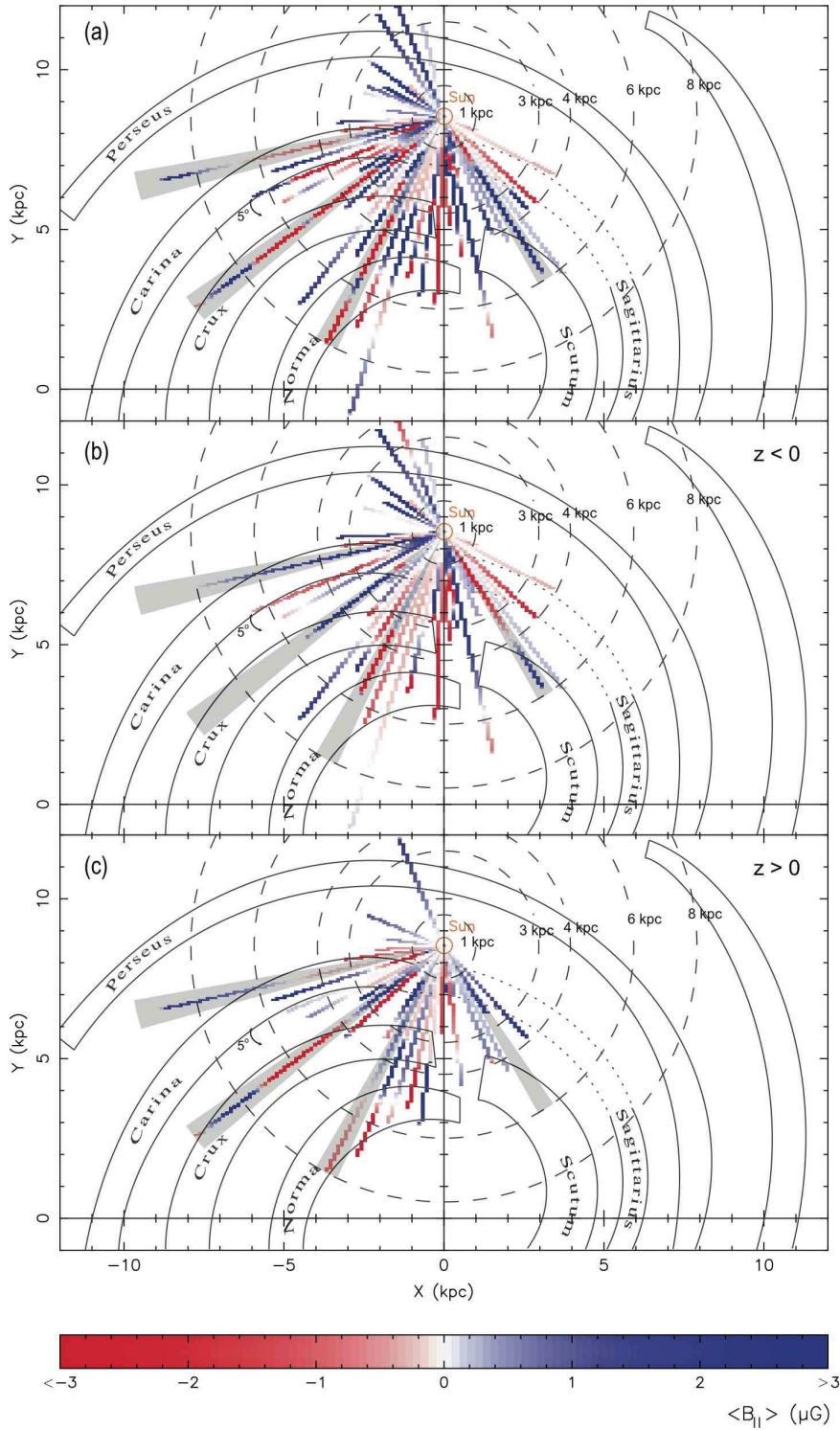} 
\caption{\label{fig:Bll2Dmap} Image maps of the projected
$\left<B_{\parallel}\right>$ values for different longitude sectors on
the GP. The field values were binned in 72, $5^\circ$ longitude
sectors and 1,000, 10-pc distance bins. Prior to calculating each
pixel's value, based on Eq.~\ref{eq:RMDM_gradients}, we performed a linear
interpolation of the available data. Hence most pixels
do not correspond to any (projected) pulsar's position. Only bins
containing a value based on the above interpolation scheme are
plotted, and thus each sector terminates at the location of the last
pulsar available in the sector. The pixel colour and intensity along
the radial paths (each representing a longitude sector of
$\delta\ell=5^\circ$) correspond to different
$\left<B_{\parallel}\right>_d$ values and are graduated from red, for
fields directed away from the observer, to blue, for fields directed towards
the observer, according to the scale shown below the plots. The
colour scale has been intentionally saturated at both ends to reveal
the field gradients in $-3\leq \left<B_{\parallel}\right> \leq 3$
$\mu$G; hence, the colour at both edges corresponds to an absolute
field magnitude of at least 3 $\mu$G. In order to reveal potential
differences in the GMF either side of the GP, we produced the maps 
for (a) all pulsars, (b) those with $z<0$ and (c) those with $z>0$.}
\end{figure*}

Fig.~\ref{fig:Bll2Dmap} shows image maps, as described above, of the
projected value of the parallel magnetic-field component binned along
72, $5^\circ$ longitude sectors and 1,000, 10-pc distance bins. The
number of pulsars per longitude sector is distributed as shown in
Fig.~\ref{fig:zldistrib}b. Only bins with available measured or
interpolated data are shown: the interpolation was only possible if a
bin lay between measured values; extrapolation was not performed. The
different pixel intensities correspond to different values of the
average field: the graduated shades of blue correspond to the range of
positive GMF, i.e.~directed towards the Sun, and the graduated shades of red
correspond to negative GMF, i.e.~directed away from the Sun. Both ends of the colour
scale represent a minimum absolute value of 3 $\mu$G --- meaning that
stronger fields are represented by the same colour. In order to test
the hypothesis of models supporting a field reversal as one crosses
the GP (i.e.~the class of models denoted with the suffix `-A'), we plotted the
image maps based on (a) all available data in each
sector, (b) only pulsars that lay below the GP and (c) only pulsars
that lay above the GP.

\subsubsection{Crux}
\begin{figure*} 
\vspace*{10pt} 
\includegraphics[width=1\textwidth]{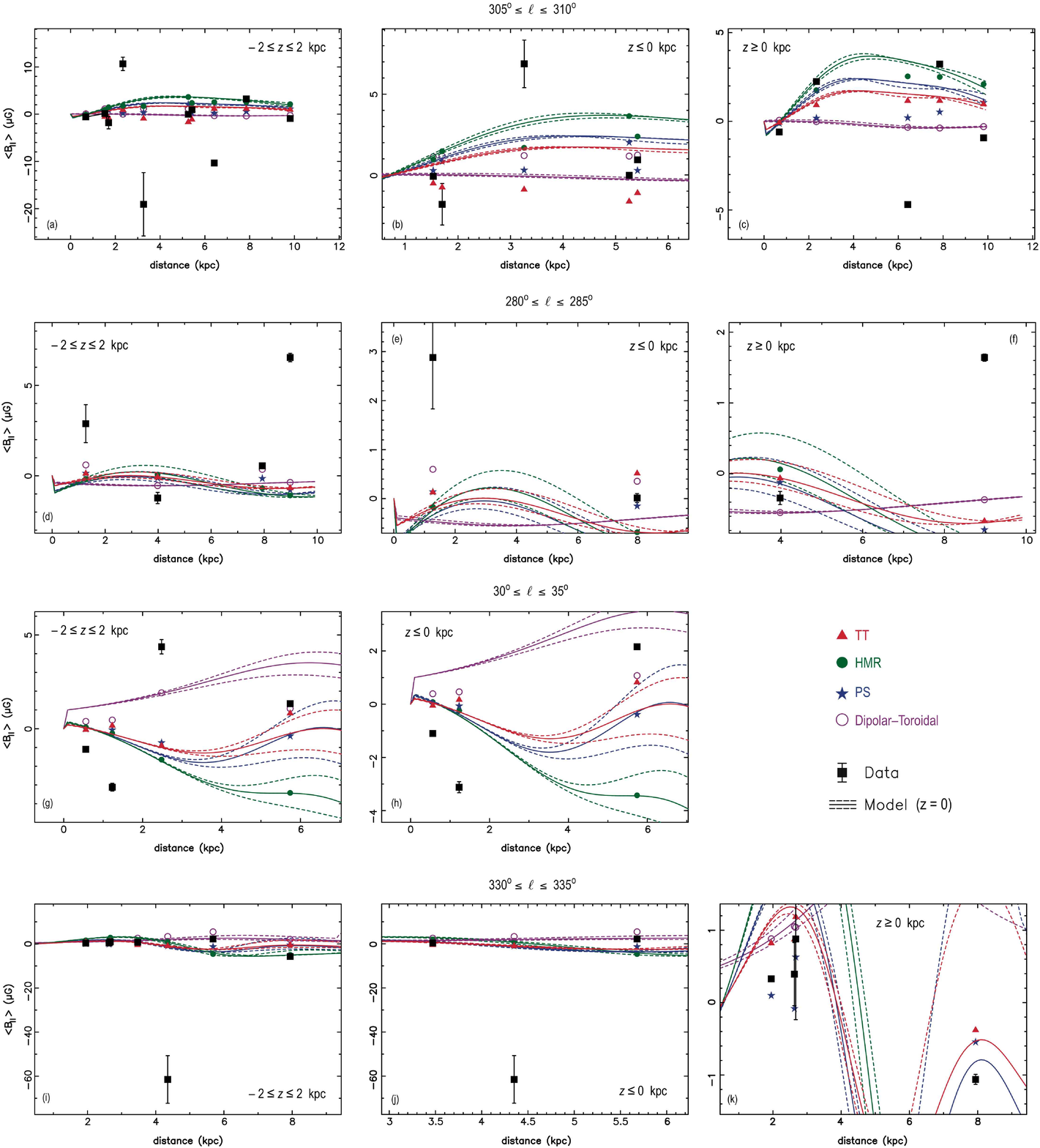} 
\caption{\label{fig:BllScatter} Scatter plots of the projected
$\left<B_{\parallel}\right>_d$ as is calculated from the gradients of
RM and DM between the available pulsars lying along three
longitude sectors: $305^\circ\leq\ell\leq 310^\circ$ (top row),
$280^\circ\leq\ell\leq 285^\circ$ (middle row) and
$30^\circ\leq\ell\leq 35^\circ$ (bottom row). For each longitude
sector, the field profile is shown based on the entire sample of
pulsars (left column), only those that lie below the GP (middle
column), and only those that lie above the GP (right column). The
projection was applied to the sample of pulsars inside $-2\leq z\leq
2$ kpc. A magnetic-field reversal in the region with $3\leq d\leq 5$
kpc, coinciding with the far edge of the Carina--Crux interarm region,
is the most likely explanation for the sudden transition from positive
(field directed towards the Sun) to negative values of the projected
field. Also plotted are the model values of
$\left<B_{\parallel}\right>_d$ from each model at each pulsar's
position: red triangles for the TT model, green circles for the HMR
model, blue stars for the PS model and purple open circles for the
Dipolar--Toroidal model.  Finally, the continuous function of
$\left<B_{\parallel}\right>_d$ with distance, for each model, is
plotted for $z=0$: the solid lines correspond to the central longitude
value of $\ell=307.5^\circ$, and the dashed lines, to the {\em edges}
of the longitude region.}
\end{figure*}

One of the most evident reversals revealed by the map of
Fig.~\ref{fig:Bll2Dmap}a coincides with the {\em los} tangential
to the Crux arm, covering 305$^\circ$--310$^\circ$ of Galactic
longitude. The scatter plot for that sector presents a complex picture
(see Fig~\ref{fig:BllScatter}a): the field is reversed from CW to CCW
somewhere between the Carina and Crux arms (2--4 kpc); an increase of
the field in the Crux arm region (5--10 kpc) is observed at larger
distances, which however fluctuates around zero on a distance-scale of
$\sim 2$ kpc. The field behaviour becomes less complex if one
separately examines the field profiles below and above the GP (see
Fig.~\ref{fig:BllScatter}b,c): in both $z>0$
and $z<0$ plots, the field direction between the Carina and Crux arms is CW; 
deep inside the Crux arm region, at $\approx 7$ kpc distance,
the plots for $z>0$ reveal a second field reversal, that
corresponds to a change in the field direction from CCW to CW. One has to bear in mind, of course,
that the field-direction changes were probed to only as good a
resolution as the data spacing allowed.

We can compare the above statements with those of \citet{bhg+07}, who
provided convincing evidence for a CCW field inside the arm region,
along $\ell\approx 307^\circ$. In general, our conclusions 
support the existence of a field with CCW direction
inside the Crux arm. If we exclude the positive value at $\approx 8$
kpc ($\approx 3$ $\mu$G), which produces a sudden reversal in the
field, then the two analyses are fully compatible with a CCW field
inside the arm. In addition, the presence of a reversal in the
interarm region is drawn from both analyses.

Finally, we used the analytical GMF functions of section
\ref{subsec:RegFieldModels} to calculate the discrete values of
$\left< B_{\parallel}\right>_d$ at each pulsar's position and compare
it with the corresponding data value. The results are plotted with
different symbols for each model in Fig.~\ref{fig:BllScatter}. In
addition, the continuous profiles of the models for $z=0$ show the
fluctuation of the field along the respective sector. In general, the
model values are not representative of the data and seem incapable of
describing the rapid fluctuations of the field. This is perhaps an
indication that the regular field component can only partly account
for the observed field, whereas an additional turbulent, small-scale
field is needed for the full description. For pulsars below the GP
(Fig.~\ref{fig:BllScatter}b) with $d<2$ and $>5$ kpc, one can note the
relatively good description of the data by the PS model.

\subsubsection{Carina}
The sector tangential to the Carina arm
($\ell=280^\circ$--$285^\circ$) is plotted in
Fig.~\ref{fig:BllScatter}d,e,f. When all pulsars in that sector are
considered, the field dips near $\approx 4$ kpc to become CCW, where
elswhere it maintains a CW direction. Based on the only two pulsars
below the GP, the field magnitude is decreased between $\approx 1$ and
8 kpc, while maintaining a positive value. This profile is consistent
with a field pattern that follows the Carina arm in a CW
direction. Surprisingly, the picture is reversed for the pair of
pulsars lying above the GP. The RM and DM gradients between
$d\approx 4$ and 9 kpc imply a field whose magnitude increases from a
very weak ($\approx -0.2$ $\mu$G) negative value, near 4 kpc, to a
much stronger ($\approx 1.7$ $\mu$G) positive value, near 9 kpc, well
beyond the Carina arm. This profile is indicative of a CCW field
reversing to CW somewhere between 4 and 9 kpc. In comparison,
\citet{hml+06} reported a CCW field in the Carina arm, but noted an
evident change in the slope of $\Delta {\rm RM}/\Delta {\rm DM}$ between 3 and 5
kpc, which they referred to as the ``Carina anomaly''. Such a
reversal, which the authors support is possibly attributed to ${\rm
H}_{\rm II}$ regions, is consistent with the profile of
Fig.~\ref{fig:BllScatter}d.

Apart perhaps from the most remote pulsar in this sector, the
comparison between models and data in Fig.~\ref{fig:BllScatter}a seems
to favour the Dipolar--Toroidal model as the best description of the
field fluctuations. Nevertheless, the model values are still more than
1 standard deviation away from the data, and furthermore the
individual plots for either side of the GP show a clearly inconsistent
picture between the models and the data points. A mitigating factor in
this case could be the existence of ${\rm H}_{\rm II}$ regions, which
--- if real --- pose a clear restriction to large-scale modelling.

\subsubsection{Scutum}
Another interesting longitude sector on the image map of
Fig.~\ref{fig:Bll2Dmap}a is the one that is tangential to the Scutum arm,
between $30^\circ$ and $35^\circ$. It is also a direction which was previously
poorly sampled and benefited mostly from the new RMs. According to
Fig.~\ref{fig:BllScatter}g, a reversal occurs between 1 and 2 kpc distance,
inside the Sagittarius arm. The reversal is located near the far edge (from the Sun) of the Sagittarius arm, 
and it changes the field direction from CW to CCW. Fig.~\ref{fig:BllScatter}g,h seem to support
a predominantly CW field in the Q1 part of the Sagittarius arm, although more pulsar RMs are
needed in that region to have a solid argument. Beyond 4 kpc, this sector represents the Scutum arm, which maintains
a positive value of the field, thus implying a field with a CCW
direction. This result is in agreement with the findings of \citet{hml+06},
who reported a CCW field across the whole of the Crux--Scutum arm. Our
analysis dealt with a very limited sample of only 5 pulsars in this sector ---
one of which lies nominally beyond 50 kpc and was therefore excluded --- and
thus the available data did not allow for a better probing of the Scutum
region.

Nearly all results from modelling this sector are incompatible with
the observed data (see Fig.~\ref{fig:BllScatter}g,h). An exception,
perhaps, are the field values from the Dipolar--Toroidal model, which
weakly appear to mirror the magnetic field trend --- especially for
$d>2$ kpc. However, the level of consistency is low and cannot justify
giving any particular model the advantage in this case.

\subsubsection{Norma}
Previous studies of the field direction in the inner Galaxy have
suggested that the innermost spiral arm, the Norma arm, has 
a CCW field across its whole extent (e.g.~\nocite{hmlq02}Han et al. 2002). As it can
be seen in the maps of Fig.~\ref{fig:Bll2Dmap}a,c, the field inside
the arm (5--8 kpc) is mainly negative (apart from a very localised
positive value around 6 kpc), which implies a CCW field. The picture
becomes more complex in the Crux--Norma interarm region (4--5 kpc),
where the field shows an abrupt, negative spike in both the plot from
the total sample and that from only the pulsars below the GP. Above
the GP, the region is poorly sampled and no safe conclusion can be
drawn.

The ``spiky'' behaviour of the field's profile, which is also seen in
the Crux sector (Fig.~\ref{fig:BllScatter}a), implies magnetic fields
that are an order of magnitude larger than the average value of the
regular Galactic field ($\sim 2$ $\mu$G). The most likely explanation
for such large values is a high RM gradient between pulsars, in
conjunction with a small change in the electron density (see
Eq.~\ref{eq:RMDM_gradients}). This is a known phenomenon, where groups
of pulsars in which the distance between pulsars is relatively short
display large variations of the RM --- and sometimes with a change
of sign. A likely reason for such fluctuations is, again, the
turbulent component of the field, which is at least as strong as the
regular field \citep{jon89}, while another possible reason is the
presence of ${\rm H}_{\rm II}$ regions, which can result in steep
electron-density gradients, thus affecting the measured RMs
\nocite{mwkj03}.

\subsubsection{Solar Neighbourhood}
\begin{figure} 
\includegraphics[width=0.47\textwidth]{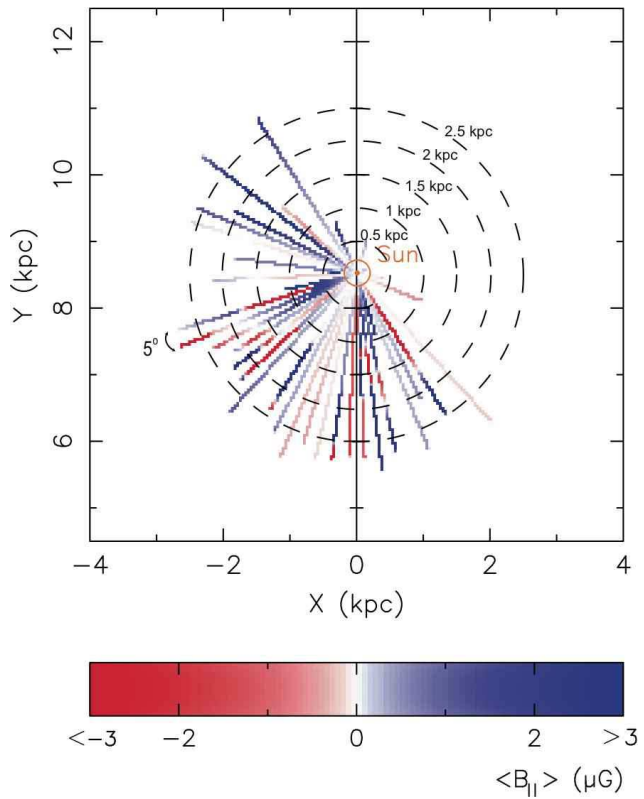} 
\caption{\label{fig:neighbourhoodMap} Image map of the projected
$\left<B_{\parallel}\right>$ values for the solar neighbourhood
($d\lesssim 2$ kpc). See Fig.~\ref{fig:Bll2Dmap} for details.}
\end{figure}

The existence of reversals within the solar neighbourhood ($d\lesssim
2$ kpc) is a common conclusion of many previous studies of the GMF
based on pulsar polarisation (see e.g.~\nocite{ls89}Lyne \& Smith 1989, \nocite{dom+01}Dickey et al. 2001,
\nocite{hml+06}Han et al. 2006). Fig.~\ref{fig:neighbourhoodMap} shows an image map,
similar to that of Fig.~\ref{fig:Bll2Dmap}a, focused on the solar
neighbourhood. At first glance, most longitude sectors include at
least one field reversal between 1 and 2.5 kpc distance. \citet{ls89}
provided evidence for a reversal between $\ell=0^\circ$ and
$70^\circ$, within $d\sim 1$ kpc. Their find can be confirmed by our
image map, which shows at least one reversal between $d=1$ and 2 kpc
in the four, best-sampled sectors of that region (see
Fig.~\ref{fig:zldistrib}b). Furthermore, there is evidence for a
similar field pattern in Q4: \citet{fsss01} performed a wavelet
analysis that suggested a reversal in Q4, between $d=0.6$ and 1
kpc. In that region, the only reversal in
Fig.~\ref{fig:neighbourhoodMap} is that between $\ell=285^\circ$ and
$290^\circ$. However, allowing for the large distance uncertainties
borne from the uneven and coarse spacing between pulsars, one could
argue that the rest of the reversals seen in that region (up to 2 kpc)
support Frick's result.

\subsection{Model Refinement via Parameter Optimisation}
\label{subsec:paramOpt}
In section \ref{subsec:locOfRevs}, it was shown that nearly all the
magnetic-field profiles along the selected longitude sectors failed to
reproduce the field variation derived from actual data. This
inconsistency could be partly attributed to the chosen values of the
different parameters that each model depends on. It has been known
that parameter optimisation can change the model predictions from
``being in dramatic disagreement with the data, to a moderately
satisfying agreement'' \citep{dom+01}.

Therefore, starting from the published parameter values for the above
four models, we applied a simple fitting routine that minimises the
reduced-$\chi^2$ between the observed and model RMs. The latter were 
calculated via path integration along the {\em los} of the parallel 
component of the model magnetic field weighted by the NE2001 density (see Appendix for more details). 
For a set of $n$ RMs and a model with two parameters,
$k_1$ and $k_2$, we can write:
 
\begin{equation} 
\label{eq:redChiSq} 
\chi^2_{n}(k_1,k_2) = \frac{1}{n}\sum_{i=1}^{n}\frac{\left[{\rm RM}_{\rm
observed}-{\rm RM}_{\rm model}(k_1,k_2)\right]^2}{\sigma_{\rm observed}^2}
\end{equation}
where our analysis considered all 150 RMs derived from the
observations.

For multiparametric models, where the number of parameters that
describe the field are $m>2$, the fitting procedure can be performed
via more than two parameters. However, the presentation of the
$\chi^2_{n}(k_1,k_2 \dots k_m)$ map poses a plotting challenge; hence
we decided to attempt to optimise only a selected pair of
parameters for each model.
 
If one takes the published parameters for each model at face value,
then the above fitting can be done through adjustment factors,
$\epsilon$ and $\zeta$, which simply translate the model-specific
parameter space to dimensionless multiples of the published values:
i.e.~the published values correspond to $\epsilon=\zeta=1$. We chose the fit
ranges for $\epsilon$ and $\zeta$ as appropriate for each
model and constructed $\chi^2_n$ maps that helped determine the optimum pair of
values for each case. The results from such fitting
procedure are described in more detail below.

\subsubsection{TT model}
\begin{figure*} 
\vspace*{10pt} 
\includegraphics[width=1\textwidth]{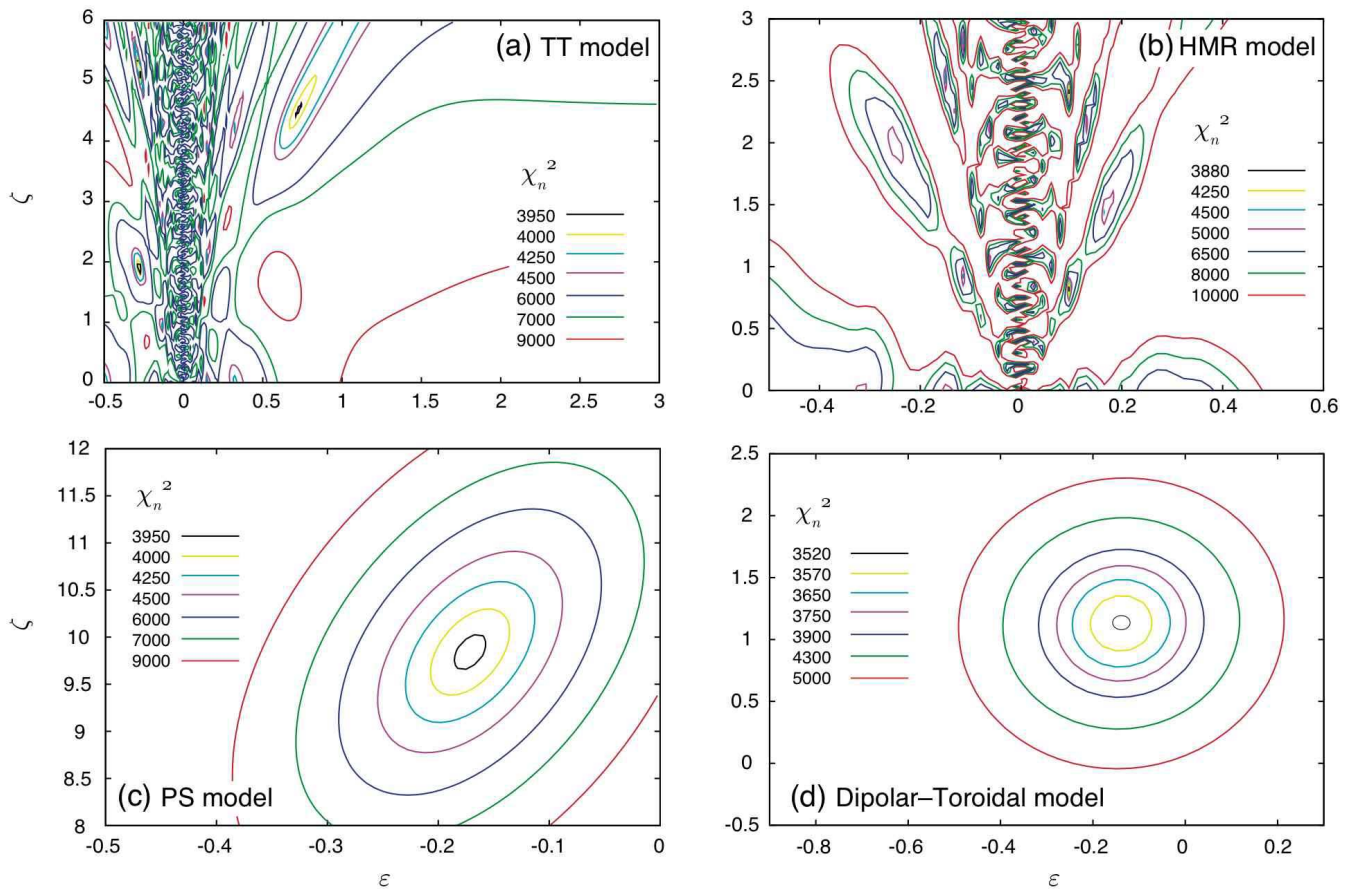} 
\caption{\label{fig:ChiSqMaps} $\chi^2_n$ maps of the adjustment
factors, $\epsilon$ and $\zeta$, from the parameter-fitting based on
(a) the TT model, (b) the HMR model, (c) the PS model and (d) the
Dipolar--Toroidal model. In all cases, $\epsilon$ and $\zeta$ multiply
the published values of the physical parameters of each model:
i.e.~(a) $p=-8^\circ\epsilon$, the pitch angle, and $d_0=-0.5\zeta$ kpc,
the distance to the nearest field reversal from the GC; (b)
$p=-10^\circ\epsilon$ and $d_0\approx-0.5023\zeta$ kpc; (c)
$b(r_\odot) = 2\epsilon$ $\mu{\rm G}$, the spiral-field magnitude at
the Sun, and $B_{T,{\rm max}}(r_\odot) = 1.5\zeta$ $\mu{\rm G}$, the
maximum toroidal-field value, 1.5 kpc above the GP, at $r_\odot$; (d)
$m=244.7\epsilon$ $\mu$G kpc$^3$, the distance-weighted toroidal-field
magnitude and $n=4.84\zeta$ $\mu$G kpc, the distance-weighted
dipole-field magnitude.}
\end{figure*}

In Tinyakov and Tkachev's model, the most obvious morphological
parameters are the pitch angle, $p$, and the distance to the nearest
field reversal to the GC, $d_0$. Our fitting approach searched for an 
optimal value in the ranges $\epsilon:[-0.5,3]$ and
$\zeta:[0,6]$, which correspond to the physical-parameter ranges
$p=-8^\circ\epsilon:[-24^\circ,4^\circ]$ and $d_0=-0.5\zeta \ {\rm
kpc}:[-3,0] \ {\rm kpc}$, respectively. The resulting $\chi^2_n$ map 
of that region is shown in Fig.~\ref{fig:ChiSqMaps}a. It is clear that
there is not a unique convergence to a minimum value of $\chi^2$ for
any particular pair ($p,d_0$). On the contrary, the map reveals a
complex dependency of the $\chi^2$ sphere on the model parameters. It
is worth noting that a number of local minima is observed, but this
cannot justify their selection as representative of optimal values, as
the former are randomly scattered across the map.

\subsubsection{HMR model}
Since the geometry of the HMR model shares many common characteristics
with that of the TT model, we chose the same parameters for the
$\chi^2$ fit: i.e. the pitch angle, which was expressed as
$p=-10^\circ\epsilon$, and the distance to the nearest field reversal,
$d_0\approx-0.5023\zeta \ {\rm kpc}$. We investigated the $\chi^2_n$ map corresponding
to a parameter space with $p:[-30^\circ,5^\circ]$ and
$d_0:[-0.5023,0]$ kpc. Fig.~\ref{fig:ChiSqMaps}b
shows a blown-up version of that map, focusing on the range where
small values of $\chi^2$ were found. Unfortunately, like in the case
of the TT model, the resulting map appears complex and without a clear
convergence. Hence, we could not optimise the HMR model's parameters via
$\chi^2$-fitting between the data and the model.

\subsubsection{PS model}
The PS model is clearly a multiparametric one, with a dipolar and a
toroidal component in addition to the disc field. Focusing on the
field magnitude instead of the morphology, we chose to try
and find the pair of values that describe best the relative magnitude
of the magnetic field in the disc compared to the toroidal one. This
decision was driven by the fact that, although there is a dipolar
field in this model, its planar component, $\mid\boldsymbol{B}_{\rm
dipole}(x,y)\mid\sim 10^{-7}$--$10^{-8}$, is 5--6 times weaker than
that of the toroidal field, $\mid\boldsymbol{B}_{\rm
toroidal}(x,y)\mid\sim 10^{-2}$; and given the fact that most pulsars
considered are very close to the GP, with more than 20\% of them with
$\mid b\mid<20^\circ$, the $B_{\parallel}$ component, which is the
quantity of interest, will mostly comprise the spiral and toroidal
components.

Our fitting process tried to determine the absolute value of the
spiral-field's magnitude at the solar radius, $b(r_\odot)$, and the
absolute value of the toroidal field maximum, 1.5 kpc above the GP, at
$r_\odot$: i.e.~the value of $B_{T,{\rm max}}(r_\odot)$. For the
spiral field, we expressed this as $b(r_\odot)=2\epsilon$ $\mu$G, and
for the toroidal field, as $B_{T,{\rm max}}(r_\odot)=1.5\zeta$
$\mu$G. The best $\chi^2_n$ value was found in the ranges
$\epsilon:[-0.5,0]$ and $\zeta:[8,12]$ or, in terms of magnetic field
magnitude, $b(r_\odot):[-1,0]$ $\mu$G and $B_{T,{\rm
max}}(r_\odot):[12,18]$ $\mu$G. Fig.~\ref{fig:ChiSqMaps}c shows a 
contour map of that region, where the minimum $\chi^2_n$ ($\approx
3685$) corresponds to $\{b(r_\odot),B_{T,{\rm
max}}(r_\odot)\}=\{-0.35,14.7\}$ $\mu$G. The negative sign of the
spiral field's magnitude expresses the fact that the field direction
is opposite to that assumed using the published values of
Eq.~\ref{eq:PStorfield}.

\subsubsection{Dipolar--Toroidal model}
In the case of a pure dipolar and toroidal field, the fitting was
performed on the respective distance-scaled field magnitudes expressed
by the parameters $m$ and $n$. Using the published values, we explored various
ranges of the quantities $m=245\epsilon$ and $n=4.8\zeta$. 
The $\chi^2$ contour map of Fig.~\ref{fig:ChiSqMaps}d shows
the region with $\epsilon:[-0.9,0.3]$ and $\zeta:[-0.5,2.5]$, where we found that
the best-fit pair of $\epsilon$ and $\zeta$ corresponds to
$\{m,n\}=\{-36.75 \ \mu{\rm G \ kpc}^3,5.37 \ \mu{\rm G \ kpc}\}$ ($\chi^2_n\approx 3516$).

\vspace*{0.5cm}

\begin{figure*} 
\vspace*{10pt} 
\includegraphics[width=1\textwidth]{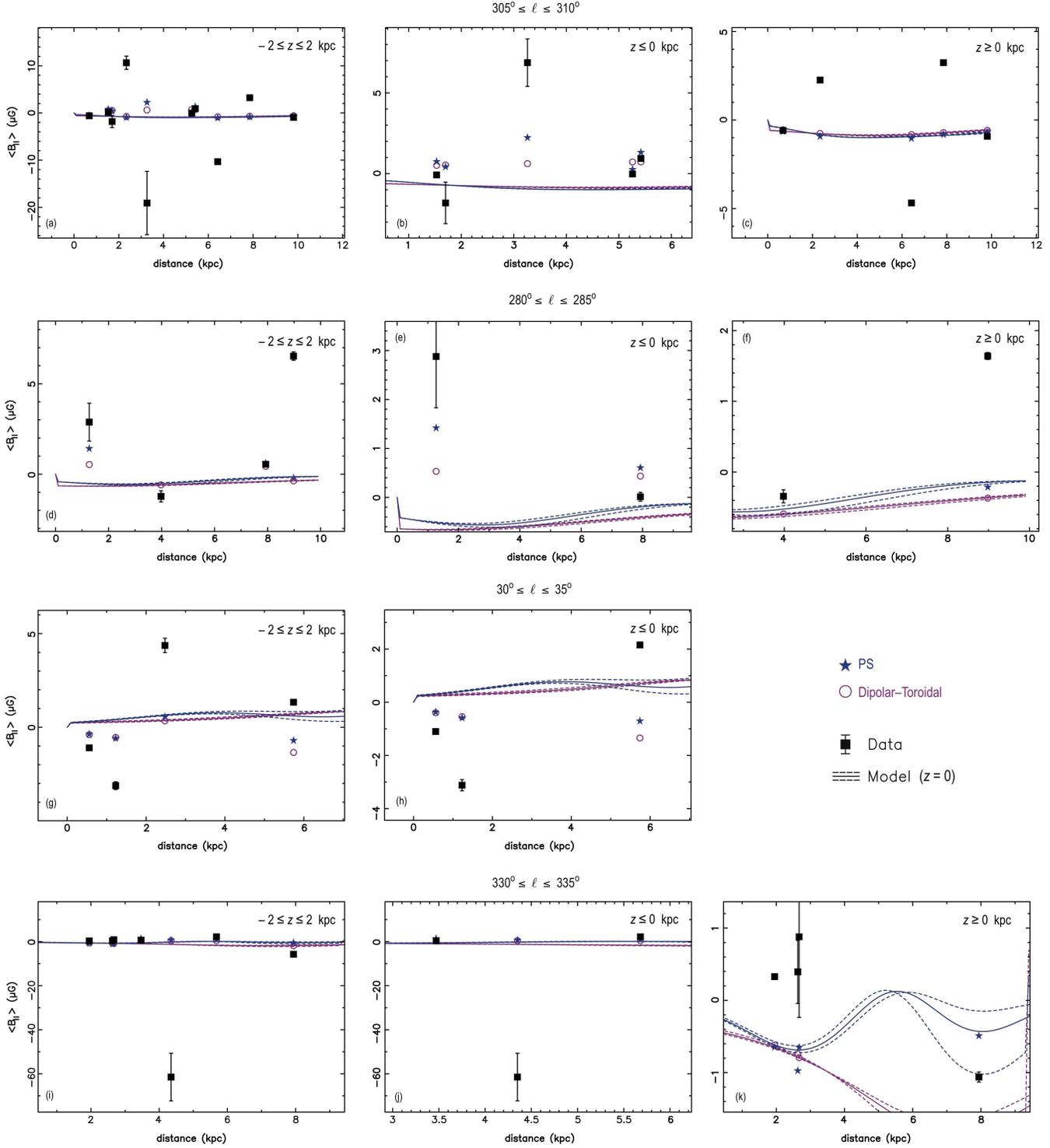} 
\caption{\label{fig:BllScatterOpt} Scatter plots of the projected $\left<B_{\parallel}\right>_d$ as in Fig.~\ref{fig:BllScatter}. The model parameters were adjusted according to the fitting procedure described in section \ref{subsec:paramOpt}.}
\end{figure*}

\noindent
Table \ref{tab:fittingResults} summarises the results from the
successful fits of the PS and the Dipolar--Toroidal models to the
data. Using the fitted values, we replotted the model predictions
for the selected regions discussed in section
\ref{subsec:locOfRevs}. The results are shown in
Fig.~\ref{fig:BllScatterOpt}.

\begin{table}
\caption{\label{tab:fittingResults} The best parameter values that
resulted from fitting model RMs from the PS and the
Dipolar--Toroidal GMF models to $n=150$ observed RMs. Each pair of
model parameters corresponds to the minimum $\chi^2_n$ found in a scan
across the respective parameter range. The negative sign in the field
magnitude implies that the field has an opposite direction to that
described by the published model parameters.}
\begin{tabular*}{0.47\textwidth}{@{}llrrrr} 
\hline
                           &                                     &  \multicolumn{2}{c}{\bf PS}     &  \multicolumn{2}{c}{\bf Dipol.--Toroidal}  \\
                           &                                     &  Pub. &  Opt. & Pub. &  Opt.         \\
\hline
$b(r_\odot)$               & \hspace{-0.3cm} [$\mu$G]            &  2         &   $-$0.35    &           &                      \\
$B_{T,{\rm max}}(r_\odot)$ & \hspace{-0.3cm} [$\mu$G]            &  1.5       &   14.7    &           &                      \\    
$m$                        & \hspace{-0.3cm} [$\mu$G kpc$^3$]    &          &         &   245       & $-$36.75                 \\
$n$                        & \hspace{-0.3cm} [$\mu$G kpc]        &          &         &   4.8       & 5.37                   \\
$\chi^2_n$                 &                                     &          &   3685    &           &   3516                 \\
\hline
\end{tabular*}
\end{table}

The post-fit plots show a marginal improvement on the data--model
consistency. In particular, the results from the PS model appear to
follow the field gradients better, although this is not the case in
terms of the absolute field magnitudes. In the region of the Carina
arm, for example (Fig.~\ref{fig:BllScatterOpt}d), the post-fit results
show better agreement between the model and data points, when compared
to the respective pre-fit plots. On the whole, between the two models
for which the fits converged, the PS model gives the most consistent
picture.

\section{Summary \& Discussion}

We used our newly determined RMs to examine particular directions in the
Galaxy where magnetic field reversal have been suggested. We did this by
using our measurements to determine the radial profiles of the average value
of the parallel component of the field along specific longitude sectors: the
selected directions were those tangential to the Crux, the Carina, the Scutum
and the Norma arms, and in addition an investigation of all directions within
2 kpc of the Sun were also examined. Furthermore, we presented an overall picture of the
magnetic field's direction by means of projecting the locations
of pulsars that lay close to the Galactic plane and plotting a colour-coded
image map of the field direction and magnitude along $5^\circ$-longitude
sectors.

Our work confirmed the existence of a field reversal in the
Carina--Crux interarm region ($d=5$--10 kpc), which has been derived from the
work of \citet{bhg+07}, based on extragalactic sources and pulsars. Also, a
predominantly CCW field was derived from both analyses, although our work
found an additional reversal near 8 kpc, deep into the Crux arm.

Furthermore, we concluded that the Carina arm maintains a CW field with
the exception of a negative value at 4 kpc, which reverses the field to
CCW. If this field reversal is real, it confirms previous claims, by
\citet{hml+06}, of a ``Carina anomaly'' somewhere between 3 and 5 kpc.

Despite the poor sampling of the Galactic region tangential to the
Scutum arm, beyond $d=5$ kpc, we confirmed the existence of a CCW field, as has
been previously reported. The only pulsar available in
the Sagittarius--Scutum interarm region suggests that the field
direction is retained there as CCW, whereas a field reversal, at
smaller distances, changes the field in the Sagittarius arm to CW.

Previous studies have indicated that the Norma arm may possess a CCW field. If this is indeed true,
our findings are compatible with such field. However, it is worth noting
that we derived a CCW direction from only two pulsars with known RMs inside the arm itself.

In many of the scatter plots of the magnetic field against distance,
there exist strong spikes where the field values are many times that
of the expected average. This displays the vulnerability of this
method to samples with neighbouring pulsars having completely
different RMs.

Finally, an examination of RMs in the local neighbourhood ($d<2$
kpc) revealed a number of reversals in Q1 and Q4. Although this
reinforces the evidence for such reversals within 2 kpc, the sparse
sampling prevented our analysis from determining the locations
of the reversals more accurately; and, hence, it was unable to confirm
results from previous, more focused studies on the region within a kpc
from the Sun.

In an effort to use the new and revised RMs as a test bench for large-scale
GMF modelling, we implemented a simple fitting routine. Using a selected
number of existing model descriptions and the NE2001 free-electron density
model, we calculated the RM predicted by these models for each of the 150
pulsars by means of integrating the electron-density-weighted model magnetic
field along the line of sight to the pulsar. We found that the overall
agreement of model RMs and observed data is generally poor. Even our 
attempts to tune key parameters of each model by a $\chi^2$-minimisation 
algorithm were only rewarded by limited success.

The difficulty to fit even a small sample of well determined RMs suggests two
possibilties. On one hand, it is possible that none of the studied models
actually represents the true magnetic field structure. On the other hand,
small scale variations in the ISM and the GMF may alter the observed RMs
sufficiently enough to cause the seen deviations from the model
expectations. The success of the models to explain some of the discussed
features suggest that the latter is the case. The aim must be therefore,
as attempted by other authors before (e.g.~Brown et al.~2007), to combine
a sample of RMs as large as possible. This sample should consist of both pulsar
RMs as well as those of extragalactic sources to provide calibration on the
largest distances. Even then, the information may not be sufficient, and
additional constraints for fitting the models should be obtained from 
observations other than RM measurements (see~e.g.~recent work by Sun et 
al.~2007\nocite{srwe07}). We will revisit such an approach in more detail
in a forthcoming publication.

\section{Conclusions}

We used a novel technique to accurately determine rotation measures from 150
pulsars, including 46 new and 104 previously published rotation measures. Based on these
values, we made an effort to investigate --- and either confirm or reject ---
previous results on the direction of the regular Galactic magnetic field at selected regions.

We checked four models of the regular Galactic magnetic field for consistency with the
data: the Tinyakov--Tkachev, the Harari--Mollerach--Roulet, the Prouza--{\v S}m{\'{\i}}da 
and the Dipolar--Toroidal model. The
magnetic-field values derived from all the models show large
deviations from the data. After optimising the Prouza--{\v S}m{\'{\i}}da and
Dipolar--Toroidal models, for which the $\chi^2$ fitting algorithm
converged to a unique value, there was a slight improvement on the
agreement between data and models. The model which benefitted mostly
from the optimisation was the Prouza--{\v S}m{\'{\i}}da model. However, the
magnetic-field-magnitude inconsistencies remained at large. A probable
explanation for these inconsistencies is that the models do not
provide the full description of the Galactic magnetic field and that a more complete
model, which will include the short-scale, turbulent Galactic-magnetic-field component, is
needed. As well as lacking a short-scale component, our analysis
uncovered the difficulties intertwined with the effort to fit a global
magnetic field model to the data. It is almost certain that our work will be followed up with more 
extensive data sets and more exhaustive algorithms. However, it remains to be seen whether
a successful model of the entire Galactic magnetic field is realisable. 
Perhaps we will have to wait until the era of the Square Kilometre Array,
with its ability to obtain rotation measures for a huge number of sources,
before significant further progress can be made.

\bibliography{journals,modrefs,psrrefs,crossrefs}

\vfill

\pagebreak

\clearpage

\renewcommand{\theequation}{A-\arabic{equation}}
\setcounter{equation}{0}  
\section*{APPENDIX}  

\begin{figure} 
\vspace*{10pt} 
\includegraphics[width=0.47\textwidth]{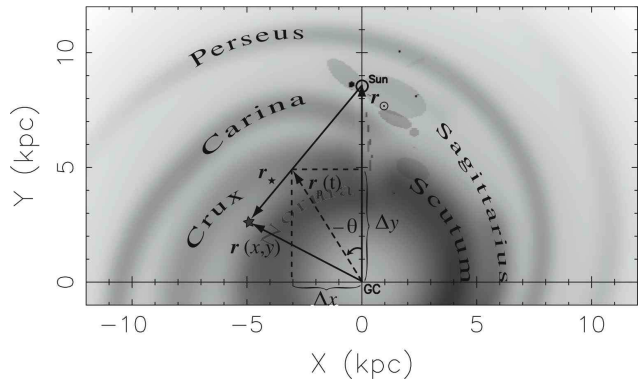} 
\caption{\label{fig:losvectors} Schematic of the position vectors that
are used in the path integration along each pulsar's {\em los}
for the calculation of the model RMs. The coordinate system used here 
matches that of the NE2001 model: a right-handed Galactocentric Cartesian system, originating at the GC, 
and with the positive $y$-axis passing through the GC and the Sun. The azimuthal angle, $\theta$, 
in this system increases clockwise from the GC--Sun direction.}
\end{figure}

\citet{cl01} provide a FORTRAN routine to calculate the electron
density, based on their NE2001 model, given a position ($x$,$y$,$z$)
in Galactocentric coordinates. The DM calulation, when using this
routine, is performed by integration of the electron density function
along the {\em los} to a particular pulsar. Since the pulsar position,
the Sun (reference) and the density function are expressed in the
Galactocentric Cartesian coordinate system, the {\em los} has to be
expressed in the same reference frame. So, the position vector
$\boldsymbol{r}_p$ to any point along the {\em los} to the pulsar is
parametrised as
$\boldsymbol{r}_p(u)=\boldsymbol{r}_\odot+(\boldsymbol{r}-\boldsymbol{r}_\odot)u$
(see Fig.~\ref{fig:losvectors}). The integration path along the {\em
los} is, therefore, expressed as $\boldsymbol{r}_{\rm
los}(u)=\boldsymbol{r}_\star
u=(\boldsymbol{r}-\boldsymbol{r}_\odot)u$.

Hence, the DM integral is expressed as

\begin{equation} 
\label{eq:DMintegral} 
{\rm DM} = \int_{0}^{1}n_e(x,y,z)\frac{dr_{\rm los}(u)}{du}du
\end{equation} 
and the path integral for the calculation of RM, as

\begin{equation} 
\begin{split}
\label{eq:RMintegral} 
{\rm RM} &= -\int_{0}^{1}n_e(x,y,z)\times \\ &\times\left[B_x\frac{dx_{\rm
los}(u)}{du}+B_y\frac{dy_{\rm los}(u)}{dt}+B_z\frac{dz_{\rm
los}(u)}{du}\right] du
\end{split}
\end{equation} 
where the adopted analytical expression of the GMF vector,
$\boldsymbol{B}(x,y,z)$, is taken from the models described
in section \ref{subsec:RegFieldModels}.

\vspace{10cm}

\end{document}